\documentclass[a4paper,fleqn,usenatbib]{mnras}

\usepackage{txfonts}

\usepackage[T1]{fontenc}
\usepackage{ae,aecompl}

\usepackage{graphicx}	
\usepackage{longtable}
\usepackage{pdflscape}

\def\Planck{\textit{Planck}}
\def\chandra{\textit{Chandra}}
\def\xmm{\textit{XMM-Newton}}

\title[The cool core state of \Planck\ SZ-selected clusters]{The cool core state of \Planck\ SZ-selected clusters versus X-ray selected samples: evidence for cool core bias.}

\author[M. Rossetti et al.]{M. Rossetti$^{1,2}$\thanks{E-mail: rossetti@iasf-milano.inaf.it},
          F. Gastaldello$^{1}$, D. Eckert$^3$, M. Della Torre$^2$, G. Pantiri$^2$, 
         \and
          P. Cazzoletti$^{2,4}$, S. Molendi$^{1}$\\
$^{1}${INAF, Istituto di Astrofisica Spaziale e Fisica Cosmica, via Bassini 15,
20133 Milano, Italy}\\
$^{2}${Dipartimento di Fisica, Universit\`a degli Studi di Milano,
  via Celoria 16, 20133, Milano, Italy} \\
$^3$ {University of Geneva, Department of Astronomy, 16, Ch. d'Ecogia, 1290, Versoix, Switzerland}\\
$^4${Max Planck Institute for Extraterrestrial Physics, Giessenbachstra\ss e, 85748 Garching, Germany}
}

\date{Accepted XXX. Received YYY; in original form ZZZ}

\pubyear{2016}

\begin{document}
\label{firstpage}
\pagerange{\pageref{firstpage}--\pageref{lastpage}}
\maketitle

\begin{abstract}
We characterized the population of galaxy clusters detected with the Sunyaev-Zeldovich (SZ) effect with \Planck\, by measuring the cool core state of the objects in a well-defined subsample of the \Planck\ SZ catalogue. We used as indicator the concentration parameter \citep{santos08}. The fraction of cool core clusters is $29 \pm 4$ per cent and does not show significant indications of evolution in the redshift range covered by our sample. We compare the distribution of the concentration parameter in the \Planck\ sample with the one of the X-ray selected sample MACS\citep{mann_ebe}: the distributions are significantly different and the cool core fraction in MACS is much higher ($59 \pm 5$ per cent) than in \Planck. Since X-ray selected samples are known to be biased towards cool cores due to the presence of their prominent surface brightness peak, we simulated the impact of the ``cool core bias'' following  \citet{eckert11}. We found that this bias plays a large role in the difference between the fractions of cool cores in the two samples. We examined other selection effects that could in principle affect SZ-surveys against cool cores but we found that their impact is not sufficient to explain the difference between \Planck\ and MACS. The population of X-ray underluminous objects, which are found in SZ-surveys but missing in X-ray samples  \citep{PSZ2}, could possibly contribute to the difference, as we found most of them to be non cool cores, but this hypothesis deserves further investigation. 
\end{abstract}




\section{Introduction}
\label{sec:intro}

It is often difficult to derive the statistical properties of a population of celestial sources from an observed sample which is a particular realization of the underlying population. Indeed, one has to be sure that the sample under analysis is representative and unbiased with respect to selection effects, i.e. that the method that was used to detect objects, and eventually to further select them, does not influence the properties that we want to analyze. Galaxy clusters are no exception to this rule. \\
Ever since the beginning of X-ray astronomy, X-ray observations have provided an efficient way to detect and characterize clusters. Many clusters catalogues (i.e. REFLEX, \citealt{bori04}, NORAS \citealt{bori00}, HIFLUGCS \citealt{reiprich02}, REXCESS \citealt{bori07}, MACS \citealt{ebeling01}) have been built basing on the ROSAT All Sky Survey (RASS), which was excellent in terms of sky coverage but limited in depth.  
X-ray surveys aimed at detecting extended sources, such as galaxy clusters, may become  ``surface brightness limited'' rather than ``flux limited'' at faint fluxes \citep{rosati02,pierre16}. Indeed, realistic X-ray surveys can detect extended objects up to a ``detection radius'' where they exceed the background level. 
It is thus easier to detect a cluster with a prominent surface brightness peak than an object with a shallower profile, even if they have the same flux when integrated to a physically relevant radius (i.e. $R_{500}$), typically larger than the detection radius. This selection bias which affects X-ray surveys of galaxy clusters is also known as ``cool core bias'' \citep{eckert11} and it was early recognized in the first {\it Einstein} surveys of galaxy clusters (\citealt{pesce90} and references therein). ``Cool core'' (CC hereafter) clusters are observationally characterized by a prominent central surface brightness peak associated to a temperature decrease in the inner regions and are usually considered as relaxed objects. \citet{eckert11} have shown that their number is overestimated in X-ray selected clusters samples (HIFLUGCS) because of their prominent surface brightness peak. A further bias is due to the higher luminosity of CC clusters with respect to NCC \textbf{at a given mass}, which makes the detection rate of CC higher than fainter NCC objects, in a flux limited sample suffering from the Malmquist bias \citep{hudson10}. 
Thus the ratio between CC and NCC objects, which depends strongly on CC formation scenarios and on the models of cluster evolution, is likely over-estimated in X-ray selected samples. \\
Over the last decade, an alternative method to search for galaxy clusters has received growing attention: the Sunyaev-Zeldovich effect (SZ hereafter \citealt{sun70,sun72}), with the publications of the first large catalogues of galaxy clusters from different experiments, containing from one hundred to more than one thousand objects \citep{ACT_cat, SPT_cat, PSZ1, PSZ2}. The SZ surface brightness does not depend on the redshift of the source, allowing us in principle to detect all the clusters in the universe above a given signal, regardless of their distance, and to build virtually mass limited samples of galaxy clusters. Actually the finite spatial resolution of real instruments limits the detection of the most distant objects (especially for \Planck\ whose lowest energy channel used for SZ measurement has a beam size of 10 arcmin) but the distribution of clusters in the mass-redshift plane is definitely flatter for SZ-selected samples than for X-ray samples. \textbf{Indeed, SZ surveys have detected more than 450 clusters at $z>0.5$, significantly increasing the number of known objects in this redshift range, which was limited to a few tens of clusters in X-ray catalogues before them (75 in MCXC, \citealt{mcxc}).} \\
Simulations have shown that SZ quantities do not strongly depend on the dynamical state of the clusters \citep{motl05}, showing only a modest effect of less than 10\% due to mergers \citep{krause12,battaglia12}. This is supported observationally by the small scatter in the scaling relation between the SZ total signal $Y$ and the mass (e.g. \citealt{cosmoPSZ1} and references therein). Moreover, \citet{PSZ2} showed  with MonteCarlo simulations that the morphology of the source, which is in general more irregular and disturbed for interacting systems, has negligible impact in the  detection procedure in the \Planck\ survey. In principle, CC bias may play a role also in SZ surveys: CC clusters feature a prominent peak in the pressure profiles \citep{pip_V}, which results in an increase in the central value of the Comptonization parameter $y\propto \int Pdl$ \citep{pipino10}. However, simulations have shown this effect to be small, especially for \Planck\ whose beam size is larger than the typical cluster size and is more sensitive to the total SZ signal rather than to its central value \citep{pipino10, lin15}.\\
In a recent paper (\citealt{rossetti16}, Paper I hereafter), we showed that the dynamical state of \Planck\ SZ-selected clusters is significantly different than in X-ray surveys. We measured an indicator of dynamical activity ($D_{X.BCG}$, the projected offset between the position of the X-ray peak and the one of the brightest cluster galaxy) for a representative subsample of \Planck\ clusters \citep{cosmoPSZ1} and we compared its distribution to the one of the same indicator in several X-ray selected samples available in the literature.  The distributions are significantly different and the fraction of dynamically relaxed objects is smaller in the \Planck\ sample than in X-ray selected samples, confirming the early impression that many \Planck-selected objects are dynamically disturbed systems \citep{planck_early_IX}. 
In Paper I, we suggested that the origin of this discrepancy may be due to the CC-bias affecting X-ray surveys, since dynamically relaxed objects usually host cool cores. However, we could not verify this hypothesis as $D_{X,BCG}$ is not a direct indicator of the presence of a peaked surface brightness profile, although it shows correlations with several cool core indicators \citep{sanderson09}. 
To test this hypothesis one would need to measure the presence and strength of the surface brightness peak in large SZ and X-ray samples and compare them. A first result in this direction has been presented by \citet{semler12}, who measured the concentration parameter \citep{santos08}, a CC indicator directly related to the strength of the surface brightness peak, for a small sample of clusters detected by SPT (13 objects). They compare the distribution of this indicator in their sample with the one in the X-ray selected 400d sample \citep{burenin07} and found them to be consistent, but  given \textbf{their small number of objects} they could constrain only the fraction of CCs between 7 and 59 \%. More recently, \citet{mcdonald} measured the concentration parameter, as well as other CC indicators, for a larger SPT sample but do not directly make a comparison with X-ray samples. Although it was not the main objective of their paper, \citet{mantz15} provided a first significant result, finding that the fraction of objects with a peaked surface brightness profile is significantly higher in X-ray selected samples than in SZ samples, using SPT and a small (30 objects) subset of the early \Planck\ catalogue \citep{planck_early_cat}. 
 Conversely, the recent comparison by \citet{nurgaliev16} between the SPT SZ sample and the X-ray selected 400d catalogue \citep{burenin07} does not address directly the role of CC-bias as it is based on morphological indicators which measure the deviation from symmetry of the cluster images and thus compare the dynamical state, as we have also done in Paper I. \\
 The aim of the present paper is to directly address the origin of the discrepancy in the dynamical state that we found in Paper I and to test the hypothesis that it is due to the CC-bias. We use as CC indicator the concentration parameter \citep{santos08}, since it directly measures the strength of the SB peak. We measure it for a large sample of SZ-selected clusters drawn from the \Planck\  catalogue and consistently for the X-ray selected MACS sample \citep{mann_ebe}.
The outline of the paper is as follows.
In Sec. \ref{sec:sample} we present our samples while in Sec. \ref{sec:data_analysis} we describe the reduction and analysis of \chandra\ and \xmm\ data, that we applied to both samples. We present our results and compare the distributions in Sec. \ref{sec:results}, comparing it also with previous results and other samples available in the literature. In  Sec. \ref{sec:ccbias} we discuss the role of CC-bias, trying to reproduce our results with simulations. Finally, we discuss other possibilities in Sec. \ref{sec:discussion}. In this paper, we assume a $\Lambda$-CDM cosmology with $H_0=70\ \rm{km}\ \rm{s}^{-1} \rm{Mpc}^{-1}$, $\Omega_m=0.3$ and $\Omega_\Lambda=0.7$.  \\

\section{The samples}
\label{sec:sample}

\subsection{\Planck\ cluster sample}
\label{sec:plancksample}
The starting point of our SZ-selected sample is the \Planck\ cosmology sample (PSZ1-cosmo), which has  been used for the cosmological analysis with cluster number counts described in \citet{cosmoPSZ1}. It is a high-purity subsample built from the first release of the \Planck\ catalogue of SZ sources \citep{PSZ1}, containing all the detection with highest signal-to-noise ratio ($S/N>7$) after the  application of a mask, that excludes the galactic plane and point sources and leaves $65\%$ of the sky for the survey. It contains 189 clusters: all of them have been confirmed at other wavelengths and redshifts have been associated to each cluster. The properties of the sample and its completeness are described in detail in  \citet{cosmoPSZ1}. 
The PSZ1-cosmo sample has been almost completely followed-up in X-rays with either \chandra\ or \xmm\, and is thus the ideal starting point to measure the concentration parameter (Sec. \ref{sec:cpar}) of \Planck-selected objects. The larger and more recent second release of the Planck SZ catalogue (PSZ2, \citealt{PSZ2}) has not benefited yet of a similar follow-up campaign and the analysis of this sample would thus be strongly incomplete. \\ 
We used \chandra\ data as a reference because better suited for the measurement of concentration parameters given the excellent spatial resolution. We measured concentration parameters for 154 objects with \chandra\ at $z>0.07$, using this redshift as a lower limit to accomodate 400 kpc within the \chandra\ ACIS-I field of view. For 10 objects at $z>0.07$ we measured concentration parameters with the \xmm\ data, as \chandra\ data were not available in the archive. For 5 objects in the redshift range $0.03-0.07$ we used \xmm\ data, exploiting its larger field of view to cover the cluster region used in the definition of the concentration parameter (Sec. \ref{sec:cpar}).  The remaining objects for which observations are potentially available but not used here are: 8 clusters with \chandra\ data planned or still proprietary as of July 2016, 4 clusters with \xmm\ data at $z<0.03$ (not completely covered even with \xmm) and 4 clusters at $z>0.35$, for which the core region used in our indicator is not resolved with \xmm\  ( Sec. \ref{sec:cpar}).\\
Our final sample is thus composed of 169 clusters in the redshift range $0.04-0.87$ with a median $z=0.18$ and in the mass range $(2-12)\times 10^{14} M_\odot$ (median $M_{500}=6.2\times 10  ^{14} M_\odot$).

\subsection{MACS sample}
\label{sec:macssample}
In Paper I, we compared the distribution of our dynamical indicator in the \Planck\ sample with the one of three X-ray selected samples (HIFLUGCS, REXCESS and MACS). We showed that MACS is the most suited for the comparison with \Planck\  among those samples, since its redshift and mass distributions are more similar to the ones of the PSZ1-cosmo sample. Actually, the sample that we used in Paper I is not the original MACS sample (whose selection criteria are described in \citealt{ebeling01}) but its extended version described in \citealt{mann_ebe} (ME-MACS hereafter). Both MACS and ME-MACS are drawn from the RASS Bright Source Catalogue \citep{vogesBSC}, with a flux limit $f_{RASS}[0.1-2.4\, \textrm{keV}] > 1 \times 10^{-12}\, \textrm{erg} \, \textrm{cm}^{-2} \, \textrm{s}^{-1}$.  The main difference is that MACS is limited by definition to the most distant systems ($z>0.3$), while ME-MACS extends to lower redshifts ($z>0.15$) and has an additional luminosity cut  $L_{RASS}[0.1-2.4\, \textrm{keV}]> 5 \times 10^{44}\, \textrm{erg} \, \textrm{s}^{-1}$. The ME-MACS sample is a well-defined purely X-ray selected sample, based on a flux limited survey, and it is thus well suited for a comparison between the X-ray and SZ selection. Moreover, its redshift distribution is more similar to the redshift distribution of the \Planck\ sample with respect to the original MACS and we thus decided to use it in our analysis. Finally, 104 out of the 129 clusters meeting the ME-MACS selection (listed in \citealt{mann_ebe}) have public \chandra\ observations  that we used to measure the concentration parameter (Sec. \ref{sec:cpar}).\\

\subsection{Mass and redshift distributions}	
\label{sec:massz}
\begin{figure*}
\centering
\includegraphics[width=0.31\textwidth]{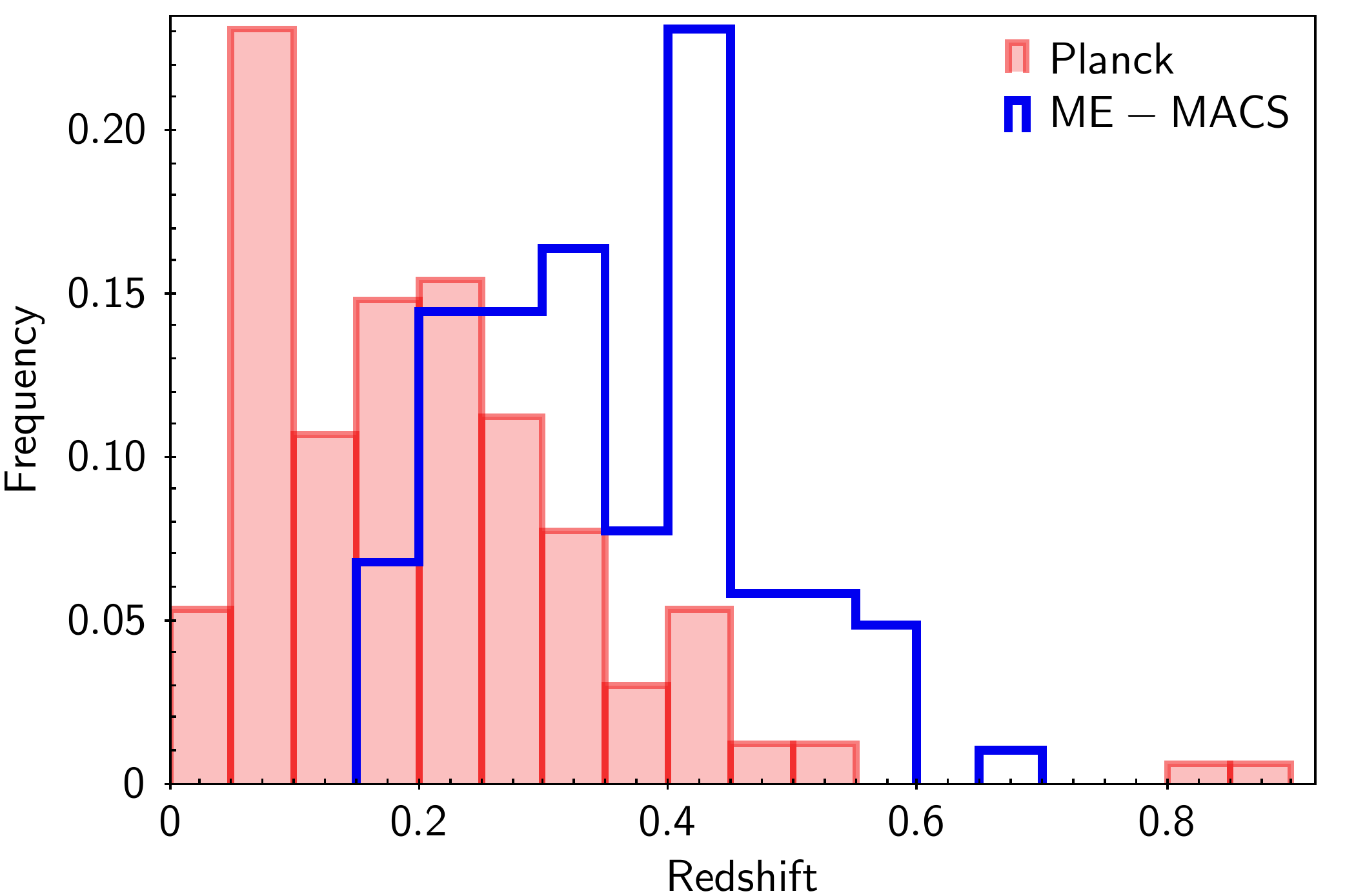}
\includegraphics[width=0.31\textwidth]{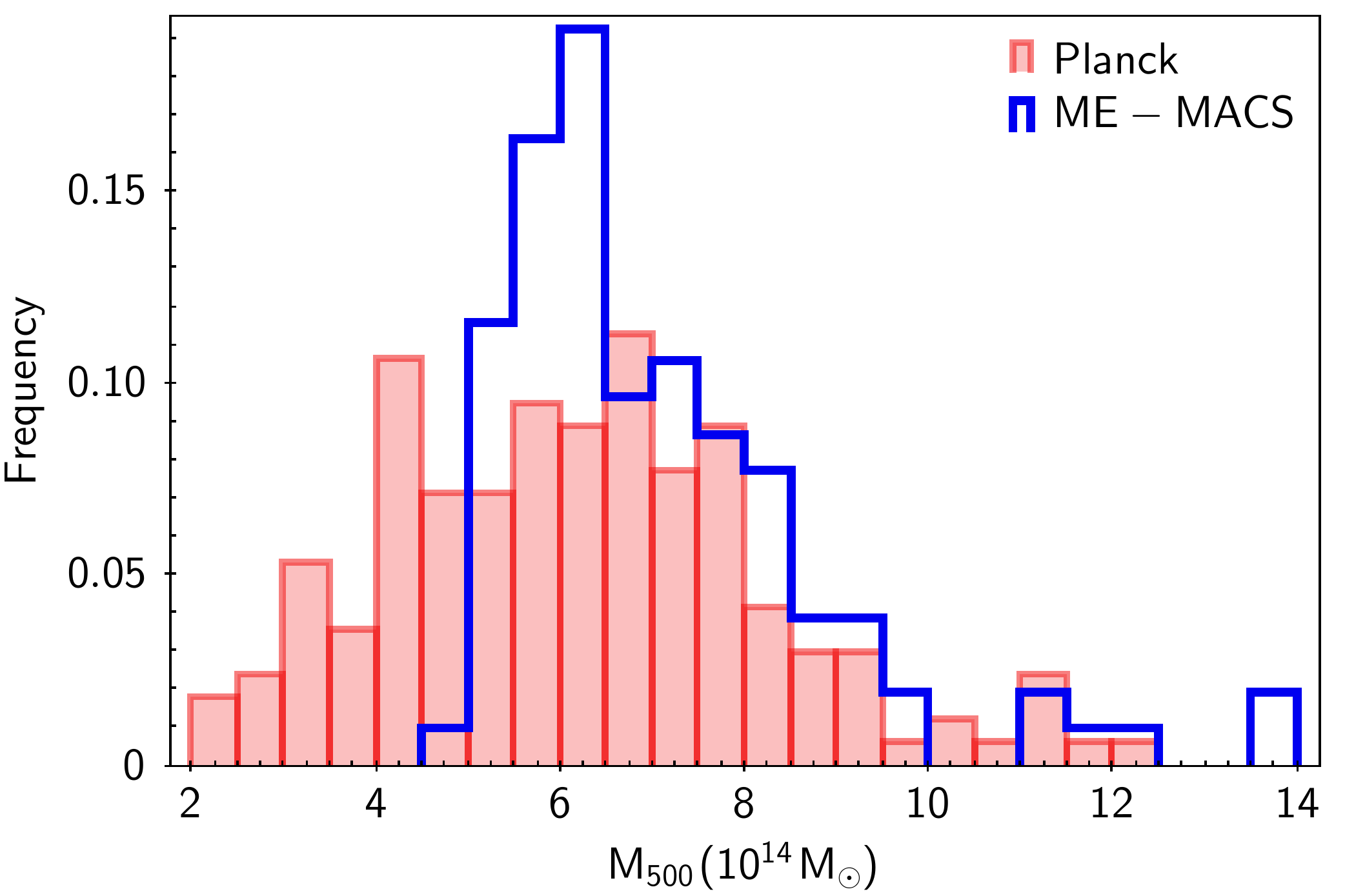}
\includegraphics[width=0.31\textwidth]{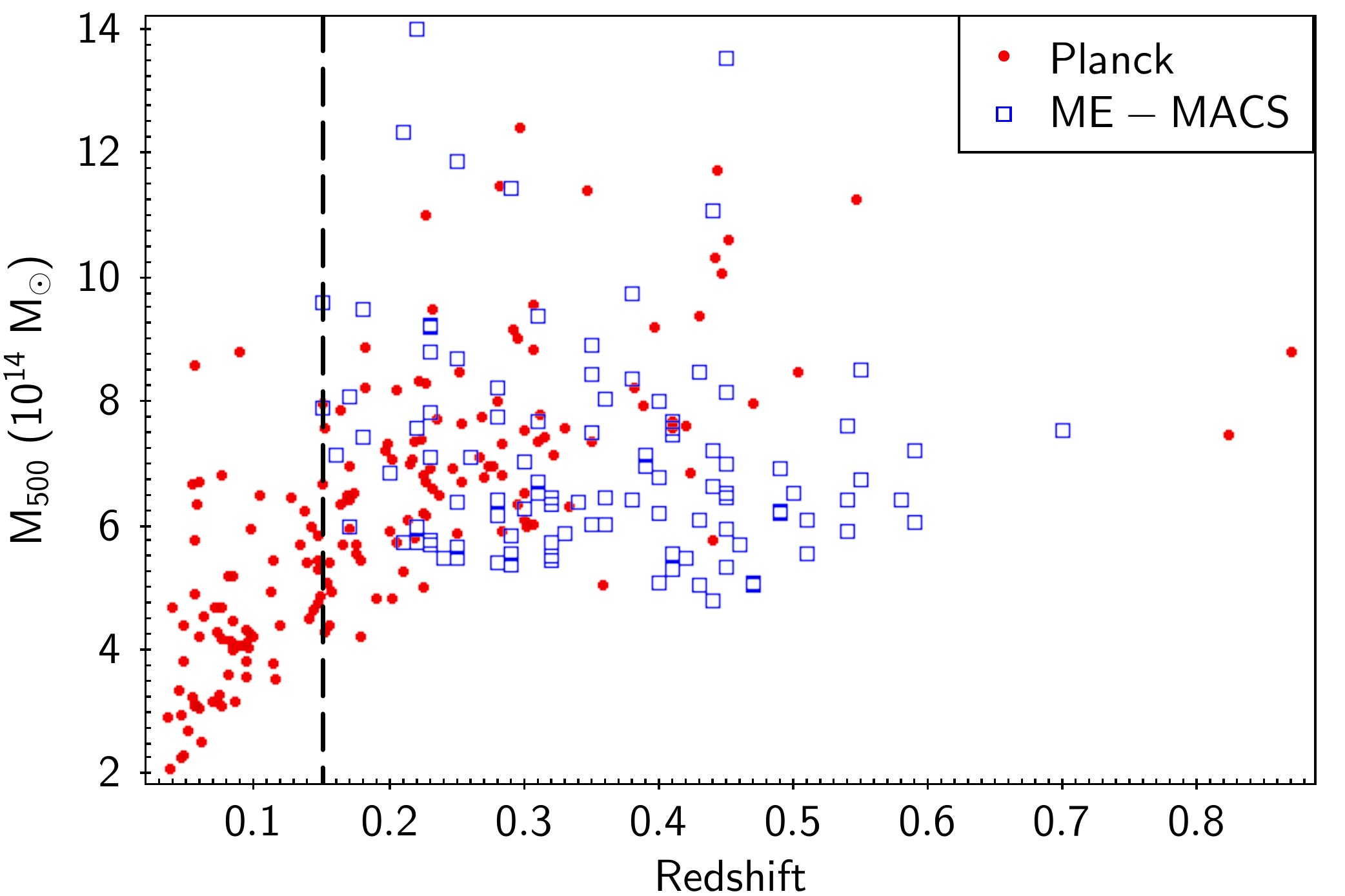}
\caption{Distribution of redshift (left), mass (middle) and $M-z$ plane (right) in the \Planck\ and ME-MACS samples.} 
\label{fig:histo_massz}
\end{figure*}

In Fig. \ref{fig:histo_massz}, we show the redshift and mass distribution of the \Planck\ and ME-MACS samples. As in Paper I, we estimate the masses of the ME-MACS clusters using the $L-M$ scaling relation in \citet{pratt09}.
By construction, the minimum redshift of the ME-MACS sample is $0.15$, thus the median value of the ME-MACS sample ($z=0.35$) is larger than in the PSZ1 sample and the two distributions are significantly different.  Also the mass distributions appear different: objects with $M<5\times 10^{14} M_\odot$ are found only in the \Planck\ sample and correspond to the low-redshift objects which are missing by construction in ME-MACS. To minimize the difference in the two samples we define a subsample of the \Planck\ catalogue by imposing $z>0.15$: with this choice we have a subsample of 103 objects , with median redshift   $0.25$ and median mass $7.1\times 10^{14} M_\odot$. 

\section{Data analysis}
\label{sec:data_analysis}
\subsection{\chandra\ data reduction}
\label{sec:chandrared}
We analyzed \chandra\ data with the CIAO software $4.6$ using CALDB version $4.6.1$, reprocessing data from the level 1 event files and following the standard data reduction threads\footnote{http://cxc.harvard.edu/ciao/threads/index.html}. We reprocessed event files using the {\verb chandra_repro } tool with standard corrections. We used as background files the blanck-sky fields provided in the CALDB database, that we reprocessed, reprojected and renormalized\footnote{We compute the renormalization factor as the ratio between the source and background count rate in the $9-12$ keV band.} to match observations. We detected point sources using the {\verb wavdetect } tool and we extracted a light curve excluding them to identify and remove periods of enhanced background. We used the  {\verb fluximage } tool to produce images in the $0.7-2$ keV bands and the appropriate exposure maps. We cleaned the images from point sources using {\verb dmfilth } by replacing the count rates in the point source region with the mean value in a surrounding annulus. From the rescaled background files we extract images in the same energy band and with the same size of the cluster images that we use for background subtraction. As our analysis is based on flux measurements in the soft band and in the central regions of the clusters (Sec. \ref{sec:cpar}), where the source outshines the background, possible systematics in the background renormalization and subtraction do not affect significantly our results. 

\subsection{\xmm\ data reduction}
\label{sec:xmmred}
We reduced \xmm\ observations with the SAS software $14.0$ starting from the raw files in the archive, which we reprocessed to produce calibrated event files.
We used the Extended Source Analysis Software (ESAS, \citealt{snowden08}) to filter periods affected by soft proton flares and to produce images in the $0.5-2.5$ keV band for each EPIC detector. We also computed the appropriated exposure maps and a model image of the instrumental background for each detector. 
We then combined the images with the {\verb comb } ESAS tool to produce EPIC images.

\subsection{Measuring the concentration parameter}
\label{sec:cpar}
For each cluster in the PSZ1 and ME-MACS sample with available X-ray images, we computed the concentration parameter introduced by \citet{santos08}:
\begin{equation}
c=\frac{F(r<40 \rm{kpc})}{F(r<400 \rm{kpc})},
\label{eq:defc}
\end{equation}
where $F(r<40\ \rm{kpc})$ is the flux within 40 kpc from the centre (representing the core region) and 
$F(r<400\ \rm{kpc})$ is the flux within $400$ kpc, representing the cluster emission. 
\citet{santos08} introduced this parameter to discriminate CC and NCC objects also at high redshift and using observations with poor statistics. They tuned the choice of the radii of the two regions (40 and 400 kpc) to separate more efficiently CC from NCC and to be able to compute $c$ with \chandra\ data both for their high-redshift sample ($z>0.7$) and for intermediate redshift clusters ($0.1<z<0.3$). Since the clusters in our samples span a similar redshift range, we decided to use the original definition of the parameter which is furthermore the most used in the literature.\\ 
To compute the concentration parameter as in Eq. \ref{eq:defc}, we calculate the intensity of the cluster emission using the background-subtracted and exposure corrected \chandra\ or \xmm\ images (Sec. \ref{sec:chandrared} and \ref{sec:xmmred}). We take into account the poissonian  noise in both source and background images and compute the error on the concentration parameter, which is typically of the order of $5\%$.
To define the two regions of interest, we need to fix a centre for the two circles and we decided to use the peak of the X-ray images, selected as the brightest pixels in the clean image after masking the point sources and smoothing it with a Gaussian with a FWHM of 7 arcseconds to reduce statistical fluctuations. When multiple observations are available for the same object, we estimated the peak from the mosaic image \textbf{to minimize the impact of statistical fluctuations especially for disturbed objects that do not feature a clear peak}. We measured the total number of net counts and its error within 40 kpc and 400 kpc, correcting for background, vignetting (through the exposure map) and CCD gaps when they intersect the regions of interest, and compute their ratio. 
When  multiple observations are available for the same object, we measured the concentration parameter on each observation and compute their weighted mean. We tested that this procedure provides consistent results than measuring the concentration parameter directly on the mosaic image.
We applied the same procedure for the clusters in  our PSZ1 sample and on the ME-MACS: we provide the estimated values in Table \ref{table:planck} and \ref{table:macs}. \\
\citet {santos10} show that for CC clusters the amount of K-correction is different for the inner 40 kpc, where the temperature is lower, than in the larger 400 kpc region and that this effect reduces the concentration parameter at high redshifts. They estimate this effect to reach 15\% for strong CCs at $z=1$, depending on the temperature in the inner region. As we do not have temperature profiles for all the clusters in our samples, we could not compute the correction factors directly for all objects. Nonetheless, we could estimate an upper limit to the intensity of this effect, by assuming for all CC clusters in our sample a minimum temperature of one third of the virial temperature, which we estimated with the M-T scaling relation by \citet{arnaud05}. We calculate the correction factor as described in \citet{santos10} and we find that the concentration parameters of CC clusters at a median $z=0.25$ should be lower by $\simeq 5\%$ with respect to their values at $z=0$.  Given the limited redshift range of our samples and the high temperature of the ICM for massive clusters, this correction is thus comparable or smaller than the statistical errors on the parameters. Nonetheless, comparison with samples or subsamples in different redshift ranges should be taken with caution.

\section{Results}
\label{sec:results}
\subsection{The distribution in the \Planck\ sample}
\label{sec:planck}

\begin{figure}
\centering
\includegraphics[width=0.5\textwidth]{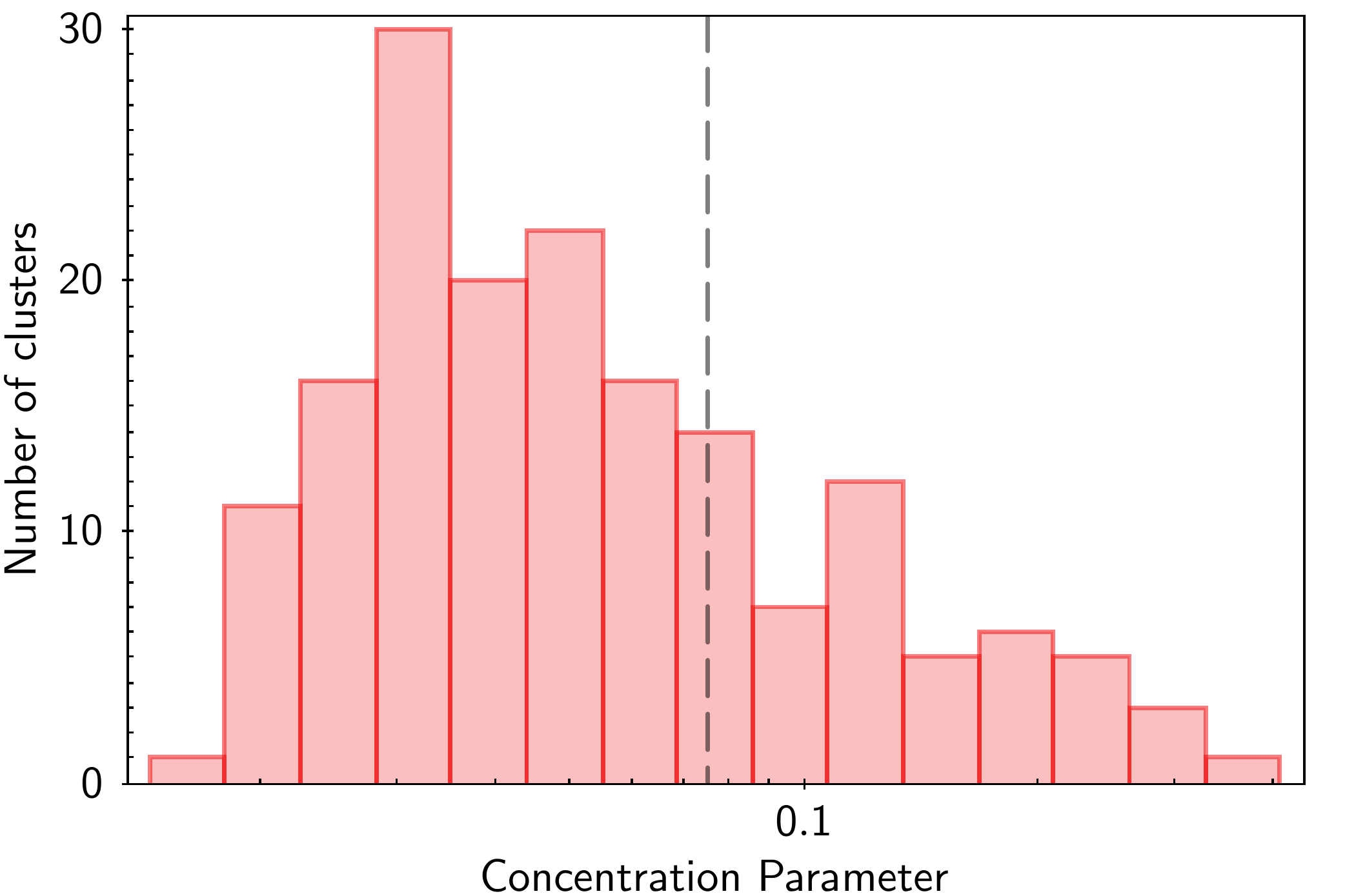}
\caption{Distribution of the concentration parameter in the \Planck\ sample. The vertical dashed line marks the threshold to separate CC ($c>0.075$) from NCC ($c<0.075$).}
\label{fig:histopl}
\end{figure}

In Fig. \ref{fig:histopl}, we show the histogram of the concentration parameter distribution in the \Planck\ sample with logarithmic bins.  The distribution features a single peak at low values of the concentration parameters with a tail extending to higher values. The median concentration parameter is $c=0.0475$.\\
We classified objects into two classes, CC and NCC, using a threshold value $c=0.075$. This value is based on \citet{santos08}, who calibrated it with the cooling time, to separate NCC and ``moderate CC'' . We merged into a single CC class the ``moderate CC'' and ``pronounced CC'' classes of \citet{santos08}. With this classification we find 49/169 CC in our sample, corresponding to a CC fraction of $(29 \pm 3)$\%, where we estimated the error with bootstrap resampling. If we consider only the subsample with $z>0.15$, we find a consistent value $(29 \pm 4)$\%.\\
As discussed in Sec. \ref{sec:plancksample}, our \Planck\ sample is not fully complete and we are missing 20 objects from the original \Planck\ cosmology sample. Even in the unlikely case that all missing clusters are classified as CC, the CC fraction of the \Planck\ sample would rise only up to 36\%. \\
\setcounter{table}{2}
\begin{table}
 \centering
\begin{tabular}{l c }
\hline
Subsample & CC fraction \\
 & \% \\
 \hline
 $z<0.18$ (median) & $27 \pm 5$\\
 $z>0.18$ (median) & $30 \pm 5$\\
 $z>0.15$ (ME-MACS) & $29 \pm 4$\\
 \hline
 $M<6.5\times 10^{14} M_\odot$ & $24 \pm 4$\\ 
 $M>6.5\times 10^{14} M_\odot$ & $34 \pm 5$\\ 
\hline
\end{tabular}
\caption{CC-fraction in redshift and mass subsamples of the \Planck\ sample.}
\label{tab:subsamplespl}
\end{table}
We divided the sample in redshift and mass subsamples to test for a possible evolution or mass dependence of the CC fraction, as measured by our indicator. The CC fractions for each subsample are shown in table \ref{tab:subsamplespl}. 
We do not find a significant variation of the CC fraction with redshift in the PSZ1 sample. However, this is not in contrast with the results found by \citet{mcdonald}, who showed a significant evolution of the CC fraction, as measured by the concentration parameter, in their sample drawn from the SPT SZ catalogue. In fact, the evolution in the SPT sample becomes evident only in redshift bins at $z>0.3$, a redshift range where we have only 24 objects in our \Planck\ sample. Indeed, at $z>0.3$, we could measure a CC fraction of 29\% with an error of 9\%,which does not allow us to draw any conclusion. Moreover, as discussed in Sec. \ref{sec:cpar} we could not apply the K-correction to our dataset and this prevents us from deriving strong constraints on the evolution of the CC fraction.\\ 
Concerning the mass dependence, we see a small difference in the CC fraction, with the low mass subsample featuring a lower CC fraction than the high mass subsample. 
Although this result is likely a statistical fluctuation ($1.5\sigma$), it is interesting to note that a similar behaviour has been found also by \citet{mantz15}: using their SPT sample they find a higher fraction of ``peaky'' objects among hotter clusters, while they do not find a similar trend for X-ray selected samples. 
The opposite trend has in fact been noted in X-ray surveys, where low-mass objects are predominantly CC (e.g. \citealt{chen07}), possibly as a consequence of the CC bias (see discussion in \citealt{eckert11}). 
Nonetheless, an increasing CC-fraction with mass is not expected and, \textbf{under the hypothesis that CCs are relaxed systems}, seems to contradict the prediction of hydrodynamical simulations that find an increasing fraction of merging clusters as a function of mass \citep{fakhouri10}.
We underline that this trend is poorly significant both in the \Planck\ and SPT sample and needs to be verified with a larger number of objects, possibly SZ selected.

\subsection{Comparison with ME-MACS}
\label{sec:compare}
\begin{figure*}
\centering
\includegraphics[width=0.49\textwidth]{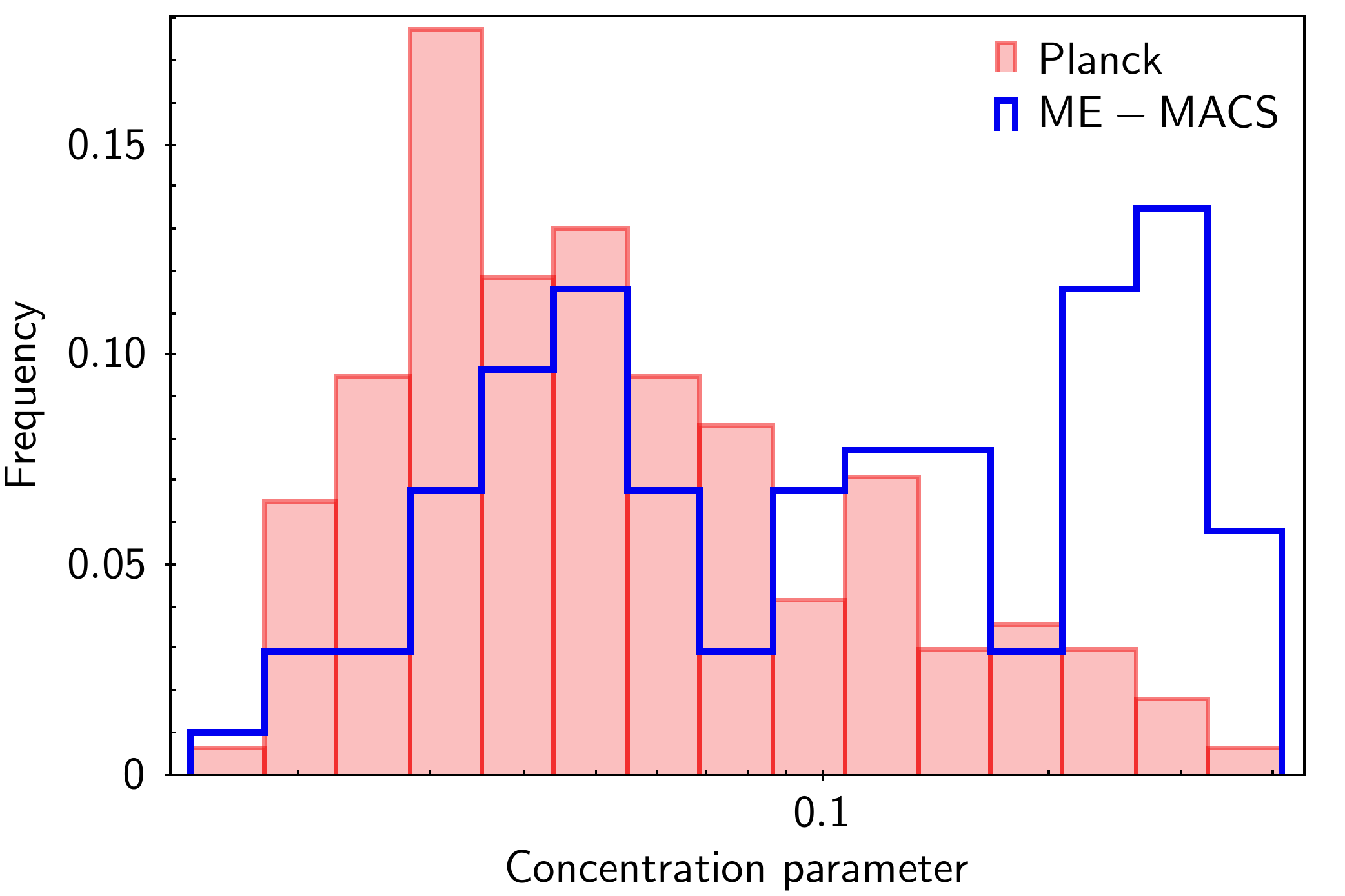}
\includegraphics[width=0.49\textwidth]{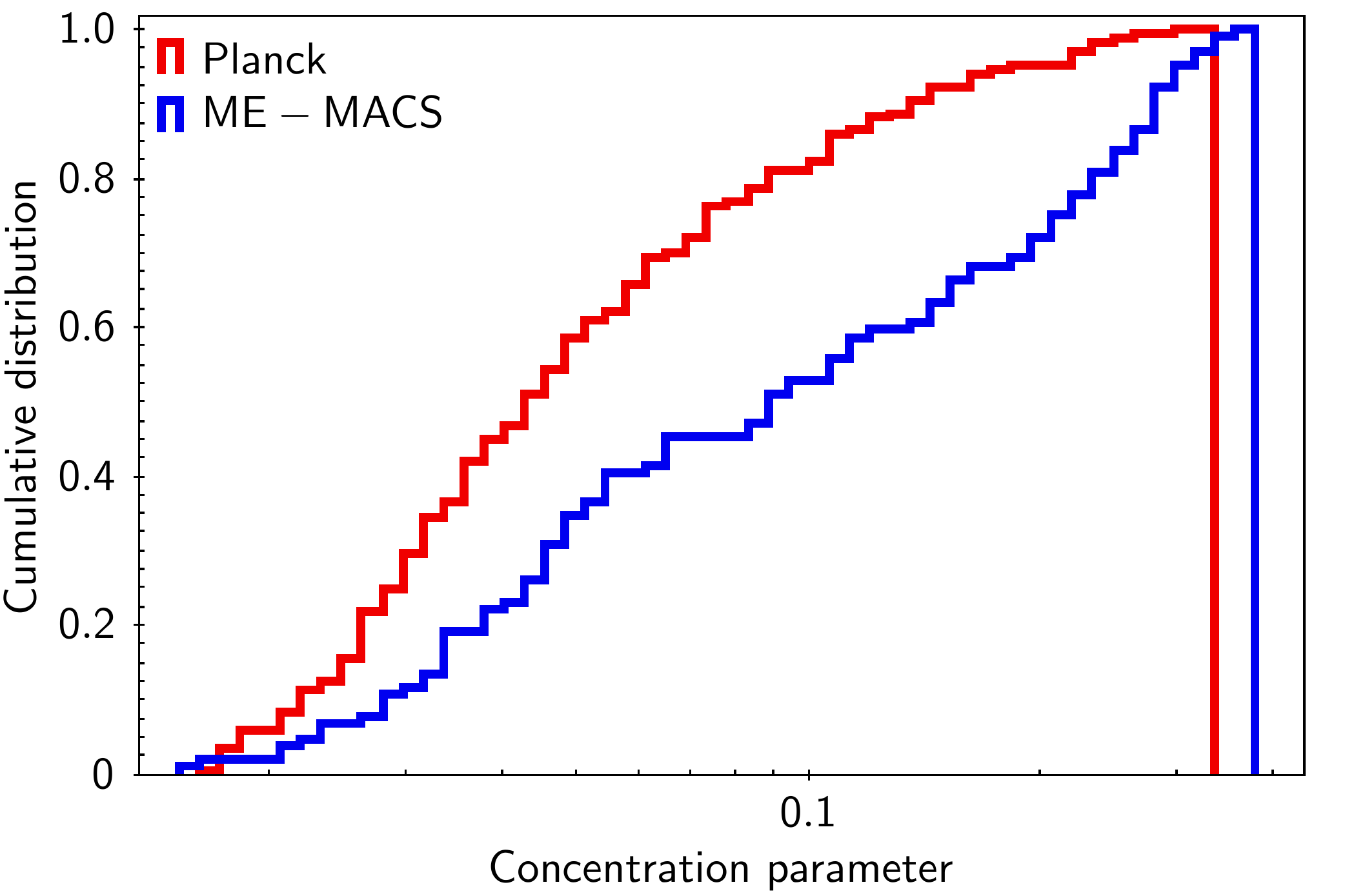}
\caption{Histogram (left) and cumulative (right) distribution of the \Planck\ (red) and ME-MACS (blue) samples.}
\label{fig:comparemacs}
\end{figure*}
\begin{figure*} 
\includegraphics[width=0.49\textwidth]{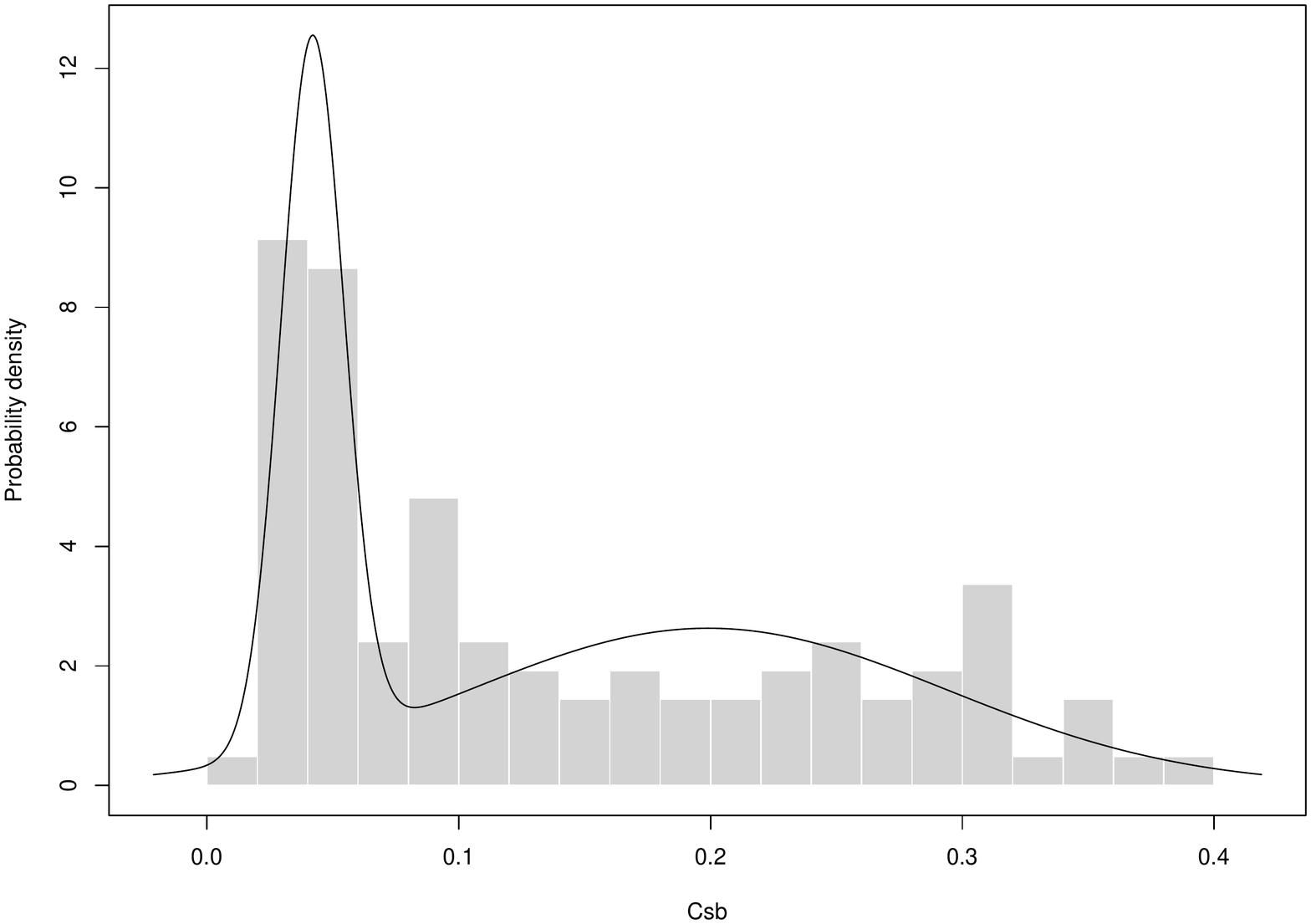}
\includegraphics[width=0.49\textwidth]{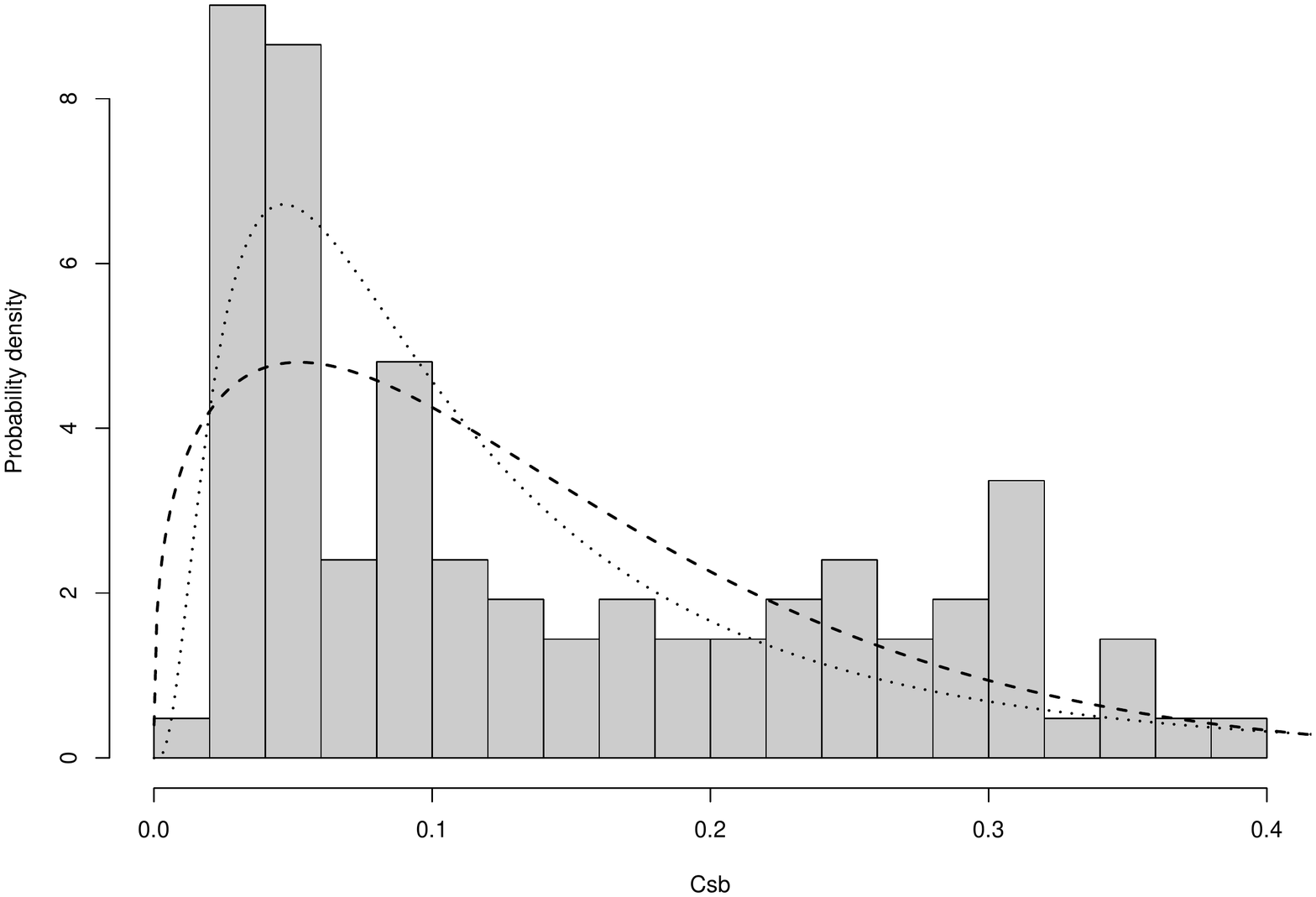} 
\caption{Left: The distribution of concentration parameters for the ME-MACS sample with the best fit two normal components model over-plotted. Right: same as in the left panel with the fit positively skewed functions: with the dotted line a log-normal and with a the dashed line the Weibull function.} 
\label{fig:macs}
\end{figure*}
\begin{figure*} 
\includegraphics[width=0.49\textwidth]{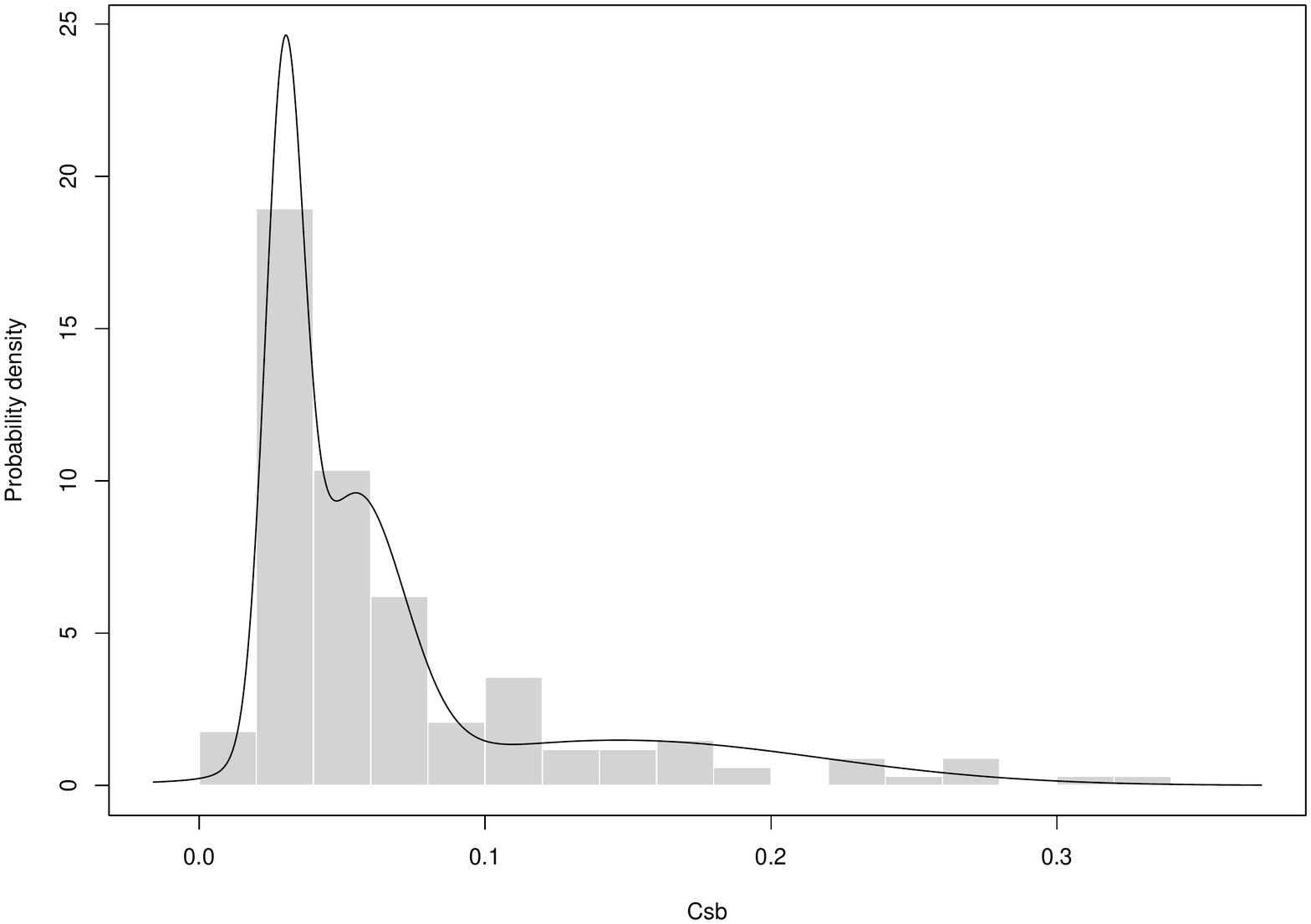}
\includegraphics[width=0.49\textwidth]{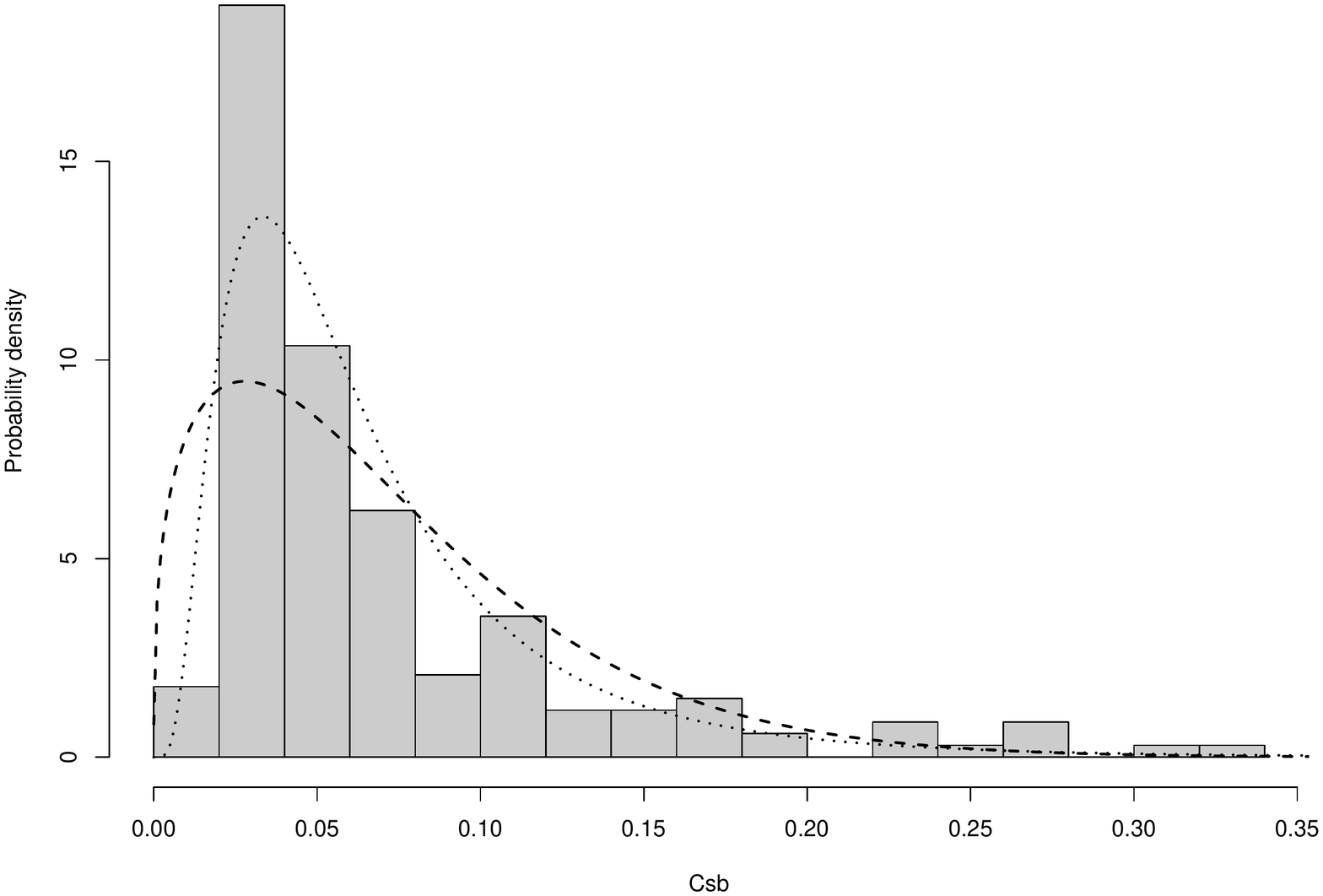} 
\caption{Left: The distribution of concentration parameters for the \Planck\ sample with the best fit three normal components model over-plotted. Right: same as in the left panel with the fit positively skewed functions: with the dotted line a log-normal and with a the dashed line the Weibull function.} 
\label{fig:planck}
\end{figure*}

In Fig. \ref{fig:comparemacs} we compare the distribution of the concentration parameter of the \Planck\ sample (described in Sec. \ref{sec:planck}) with the one in the ME-MACS sample. The distribution of the X-ray selected ME-MACS  is qualitatively different from the one of the \Planck\ sample: it shows two peaks, one for the NCC objects and one corresponding to CCs. \\
Most objects in ME-MACS are classified as CC and the CC fraction is $59 \pm 5$\%. We can compare it directly with the CC fraction in the \Planck\ sample estimated with the same indicator and the same threshold (Sec. \ref{sec:planck}): the difference is significant at more than $5\sigma$. The difference is still significant even when compared to the CC fraction in the \Planck\ subsample with $z>0.15$ ($29\pm4$\%). Even assuming that all the clusters which meet the ME-MACS criteria but do not have \chandra\ observations (Sec. \ref{sec:macssample}) are NCC, the CC fraction would decrease only to 47\%, larger at $3\sigma$ than the fraction in the \Planck\ sample.\\
We can further apply statistical tests to compare the two distributions shown in Fig. \ref{fig:comparemacs} to assess that they are different independently of the choice of the threshold separating CC from NCC. We use the Kolmogorov-Smirnov test, which measures the probability that the two samples are drawn from the same parent distribution. The KS statistic $D$, i.e. the supremum distance from the two cumulative distributions (Fig. \ref{fig:comparemacs}), is $0.349$, with a null-hypothesis-probability $p_0=1.68\times 10^{-7}$, showing that the \Planck\ and ME-MACS distributions are significantly different. If we consider only the \Planck\ subsample with $z>0.15$ (Sec. \ref{sec:massz}), the KS still returns a significant difference between the \Planck\ and ME-MACS sample ($D=0.334$ and $p_0=1.24\times 10^{-5}$).\\
The result of the KS test is supported by the qualitative difference between the two distributions: two peaks seem present in the ME-MACS sample, whereas the distribution of \Planck\ values looks more consistent with a peaked distribution with a tail at high concentration values, i.e. a positively skewed function.
We tested these differences quantitatively performing a maximum likelihood fit for each of the two 
distributions on the unbinned data using the \textsc{mclust} package \citep{mclust1,mclust2} and 
the \textsc{fitdistr} function of the package \textsc{MASS} \citep{mass} in the software 
environment \textsc{R} version 3.2.2 \citep{r_cite}.  
The model-based clustering implemented in \textsc{mclust} is an algorithm for fitting normal mixture models,
i.e. maximum likelihood fits are performed assuming a number from 1 to 9 normal components are present in
the data. The function \textsc{fitdistr} performs a maximum likelihood fit of the data to some probability
distribution functions, either calculated using analytic formulae (as for example in the log-normal case)
or computed by optimization of the likelihood. We chose for fitting two commonly used positively skewed
functions: the Weibull and log-normal distributions. We performed model selection comparing the
Bayesian Information Criterion \citep[BIC,][]{Schwarz:78} defined as ${\rm{BIC}} = 2lnL - k log(n)$ where $L$
is the likelihood, $k$ is the number of parameters of the model and $n$ is the number of data points; $klog(n)$
is the penalty term which compensates the difference in likelihood due to an increase in the number of fitting
parameters. The best model is the one that maximizes the BIC. Commonly adopted thresholds for the difference
in BIC values of two models are: a BIC difference of 0-2 is 
considered as weak evidence, 2-6 positive evidence, 6-10 strong evidence and $>10$ as very strong evidence in
favor of the model with the greater BIC value \citep{Kass.ea:95,Rafferty:95}.
We did not work on log space because the positively skewed functions are not defined for negative values.
For ME-MACS the result of the normal mixture model strongly disfavors a single Gaussian component, with
a BIC value of 159.88 with respect to two Gaussian components 
with a BIC of 217.99. A three components Gaussian model has a BIC value of
219.88 so the improvement is not significative. The two components consist of 41 and 62 members with the 
separation at a value of 0.07 (see Fig.\ref{fig:macs}, left panel), which is similar to the threshold to separate CC and NCC that we adopted in our analysis.
The fit with positively skewed functions returns BIC values of 203.70 and 202.23 for the
log-normal and Weibull distributions respectively. Those models are therefore clearly disfavored
with respect to a two components normal mixture model.
For the Planck sample the model with the highest BIC value is the one
with 3 Gaussian components, with a BIC of 610.20 which is significant with 
respect to a 2 components model which has a BIC of 597.63. The optimal
partition returns three groups with 71, 59, and 39 members respectively with
separations at values of the concentration parameters of 0.04 and 0.09 
(see Fig.\ref{fig:planck}, left panel).
The fit with a log-normal function returns a BIC value of 622.60 so this model is favored
with strong evidence with respect to the best three components normal mixture model.
A Weibull distribution is also disfavored as its BIC is 572.82. 
We can therefore conclude that the distribution of concentration parameters of the \Planck\ sample is described by a log-normal distribution, while the ME-MACS catalogue is best 
described by a bimodal behavior. The secondary peak at high concentration parameter in the latter distribution may be due to the CC bias, as the number of peaked objects is artificially boosted in X-ray surveys (Sec. \ref{sec:sim}).\\
We tested the correlation between the concentration parameter and the dynamical indicator that we used in Paper I (i.e. the projected distance between the X-ray peak and the BCG) and found them to be significantly anti-correlated in both samples. We provide the details of this analysis in Appendix A.

\subsection{Comparison with other samples in the literature}
%
\begin{table*}
 \centering
\begin{tabular}{l c  c  c c c}
\hline
Sample & CC fraction & KS test & Median $z$ & Median $M_{500}$ & Number of objects \\
 & \% & $D\ (p_0)$ &  & $10^{14}\ M_\odot$ & \\
 \hline
 \Planck\ & $29 \pm 4$ & - & $0.18$ & $6.2$ & 169 \\
 \Planck\ $z>0.15$ & $29 \pm 5$ & - & $0.25$ & $7.08$ &  103 \\
 ME-MACS & $59 \pm 5$ & $0.349$ ($1.7\times 10^{-7}$) & $0.35$ & $6.54$ & 129 \\
 \hline
 HIFLUGCS (X) & $56 \pm  6$ & $0.316$  $(1.3\times 10^{-4})$ & $0.047$ & $2.70$ & 62 \\ 
 V09 low-$z$ (X) & $58 \pm 10$ &  $0.334$  ($9.0\times 10^{-3}$) & $0.075$ & $6.18$ & 26 \\
 V09 high-$z$ /400d (X) & $31 \pm 8$ & $0.147$ ($5.5\times 10^{-1}$) & $0.49$ & $2.90$ &  36 \\  
 Pascut15 (X) & $74 \pm 5$ & $0.471$ ($1.5\times 10^{-9}$) & $0.50$ & $2.72$  & 62 \\
 Santos10 (X) & $60 \pm 13$ &  $0.421$  ($1.0 \times 10^{-2}$) & $0.82$ & $2.18$ & 15 \\
 \hline
 SPT all (SZ) & $29 \pm 5$ & $0.192$ ($3.0 \times 10^{-2}$) & $0.59$ & $5.17$ & 81 \\
 SPT low-$z$ (SZ) & $29 \pm 7$ &  $0.170$ ($2.7 \times 10^{-1}$) & $0.47$ & $5.60$ & 41\\
\hline
\end{tabular}
\caption{CC-fraction in literature samples and KS test compared with the \Planck\ sample.}
\label{tab:allsamples}
\end{table*}

As discussed in Paper I, ME-MACS is the most suited sample to be compared with \Planck\  among the well defined X-ray selected catalogues we used in that paper, and this is the reason we focused our analysis on it in the present paper. Nonetheless, the concentration parameter has been calculated for many other samples of galaxy clusters with the definition of \citet{santos08} and we can use the tabulated values for calculating their CC fraction and for doing a KS test to compare with the \Planck\ sample. We found literature information on the concentration parameter for HIFLUGCS (\citealt{hudson10}, T. Reiprich private communication), the \chandra\ Deep Group survey in \citet{pascut15}, a high-redshift sample computed by \citet{santos10} and built using WARPS \citep{perlman02,horner08} and RDCS \citep{rdcs_cat}, and the two samples used in the cosmological analysis by \citet{vikh09III}: the low-$z$ one, whose $c$ values are provided by \citealt{santos10}, and the high-$z$  subsample, drawn by 400d \citep{burenin07}, for which the $c$ values were computed by \citet{semler12}. We note that the the above samples span different masses and redshift ranges (as shown in Table \ref{tab:allsamples}) since they are derived with different limiting fluxes starting from X-ray surveys, based either on RASS (HIFLUGCS, ME-MACS) or on deep pointed PSPC observations (400d, WARPS and RDCS).
We also found tabulated value of the concentration parameter for the SZ-selected SPT XVP sample,  described in \citet{mcdonald}.The cool core fraction (using the same threshold value $c=0.075$), the results of the KS test in comparison with our \Planck\ sample as well as the median redshift and mass of each sample are provided in Table \ref{tab:allsamples}. \\
This analysis confirms that the ME-MACS sample is the most similar in terms of both mass and redshift to the \Planck\ sample. The CC fractions of most X-ray selected samples are significantly higher than in \Planck\ in all redshift and mass ranges, with the notable exception of the high-redshift sample of  \citet{vikh09III}, drawn from 400d, which features a fraction consistent with \Planck. The difference between this sample and other X-ray selected samples has been already debated in the literature (e.g. \citealt{santos10,mantz15}) and it is beyond the scope of this paper. Nonetheless, we notice that the limiting flux of 400d ($1.4 \times 10^{-13}\ \rm{erg}\ \rm{cm}^{-2}\ \rm{s}^{-1}$ \citealt{burenin07}) is higher than those of WARPS ($6.5\times 10^{-14}\ \rm{erg}\ \rm{cm}^{-2}\ \rm{s}^{-1}$) and RDCS ($1-3 \times 10^{-14}\ \rm{erg}\ \rm{cm}^{-2}\ \rm{s}^{-1}$), also based on deep ROSAT PSPC pointed observations. Possibly, the higher flux threshold imposed in 400d with respect to the detection limit reduces the effect of the CC-bias \citep{eckert11,rosati02} and allows to build a ``representative snapshot of the cluster population of typical clusters at $z=0.3-0.8$''  \citep{burenin07}. We also note that the difference in the cool core fraction of  RDCS+WARPS sample and 400d discussed by \citet{santos10} is due to the difference in limiting fluxes cited above: all the CC objects in RDCS or WARPS have a measured flux below the 400d threshold.
Given that 400d is the only X-ray sample featuring a CC fraction consistent with \Planck, it is
not surprising that \citet{semler12} and \citet{nurgaliev16} found that the distribution of concentration parameters and of morphological indicators are consistent in SPT-selected samples and 400d. However, it appears clear that 400d is rather unique among X-ray samples. \\
The only other SZ-selected sample in Table \ref{tab:allsamples} is SPT XVP \citep{mcdonald} which features a CC fraction consistent with the one in \Planck. Since the SPT-XVP sample extends to higher redshift ($0.32-1.2$) and lower masses than the \Planck\ sample, we extracted a subset from the SPT catalogue in the redshift range $0.32-0.6$ (41 objects) and compare it with the \Planck\ sample in the same redshift range (only 24 clusters). The CC fraction are in very good agreement as SPT finds $29\pm7\%$, while with the \Planck\ subsample we have $29\pm9\%$. It is certainly intriguing that both SZ-selected samples provide a similar CC fraction, but the large error bars, due to the limited number of objects in the common redshift range, do not allow us to draw strong conclusions about this agreement of the cool core fraction in different SZ-selected samples.

\section{The role of  CC-bias}
\label{sec:ccbias}
In this section, we test the hypothesis that the difference between the \Planck\ and ME-MACS distributions of concentration parameters is due to the CC-bias, first by looking at the properties of clusters which are detected only in ME-MACS and not in \Planck\ (Sec. \ref{sec:notplanck}) then by performing dedicated simulations (Sec. \ref{sec:sim}).

\subsection{Missing ME-MACS clusters in \Planck\ }
\label{sec:notplanck}
\begin{figure}
\centering
\includegraphics[width=0.45\textwidth]{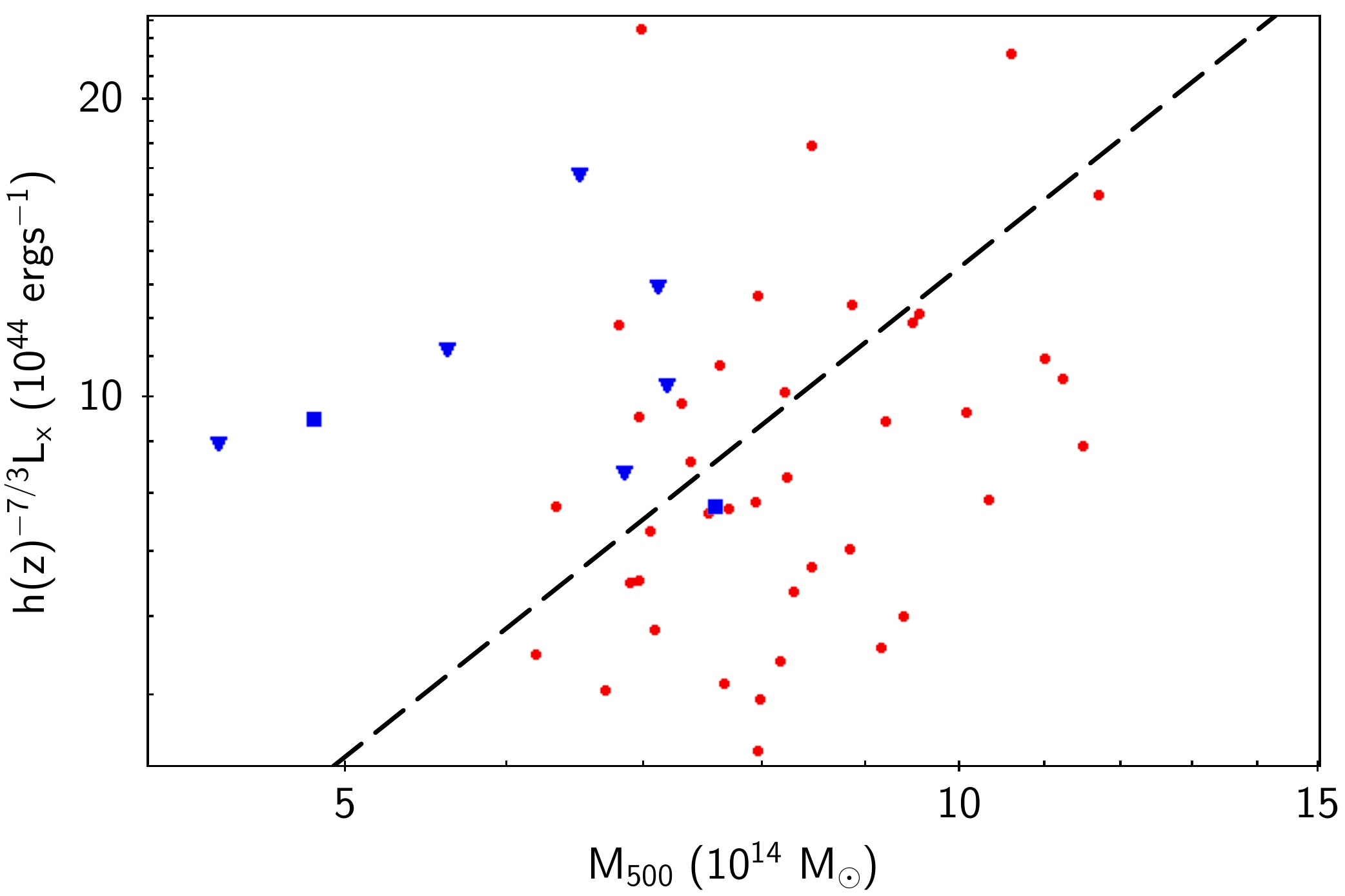}
\caption{Distribution in the $L-M$ plane of the clusters in common between \Planck\ and ME-MACS (filled red circles) and for those in ME-MACS but not in \Planck\ (blue triangles for SZ-derived mass measurement and blue square for weak lensing). The dashed line marks the scaling relation of \citet{pratt09}.}
\label{fig:notplanck}
\end{figure}

In Fig. \ref{fig:histo_massz}, we showed the distribution of \Planck\ and ME-MACS objects in the mass-redshift plane: the mass limit of the ME-MACS sample is below the one of our \Planck\ sample in the redshift range $0.2-0.6$. Therefore, we expect to find in the \Planck\ sample only the most massive ME-MACS objects, which we selected for having $M_{500} > 8\times 10^{14}\, M_{\odot}$ at $z>0.4$ or $M_{500} > 7\times 10^{14}\, M_{\odot}$ in the redshift range $0.15-0.4$. We found 36 objects in ME-MACS with the above criteria and 24 are in common with \Planck\, while 12 are not found in our \Planck\ sample.   Of these, three are located behind the galactic mask and we are thus left with nine massive objects that should be found also in the \Planck\ sample but are not. We looked at their concentration parameters and all of them are classified as cool cores. It is interesting to note that most of the ``missing clusters'' in our \Planck\ sample feature a strongly peaked SB profile ($c>0.2$), i.e. belong to the secondary peak of the ME-MACS distribution which is not found in the \Planck\ histogram (Sec. \ref{sec:compare}). \\
 We underline that the masses of ME-MACS, that we used to select potential \Planck\ clusters, are calculated from the $L-M$ scaling relation (see Sec. \ref{sec:massz}) and may thus be biased high for CCs \citep[ e.g.][]{hudson10}. For 8 of the missing objects, we found independent mass measurements, either in the \Planck\ catalogue \citep{PSZ1} (i.e. they are detected by \Planck\ with a $S/N$ in the range $4.5-7$ and thus do not enter in the cosmology sample that we analyzed here) or from weak lensing (for two of them) in the LC$^2$ catalogue \citep{sereno15}. We show their position in the $L-M$ plane along with the common objects in the two samples in Fig. \ref{fig:notplanck}. Almost all the missing objects lie above the scaling relation and their independent mass measurements are below the mass limit of our \Planck\ sample, explaining why they are not found in the \Planck\ cosmology sample. Their luminosity (and thus their mass estimate from $L-M$) is likely boosted by the presence of the cool core. \\
The fact that all the objects we considered here are CC, suggests that the CC-bias may have a role in explaining the difference between the two samples. However, to firmly test this hypothesis observationally, one would need to start from a complete population of clusters with independent mass measurements and to compare it with the properties of SZ-selected and X-ray selected samples with similar mass limits, to see which clusters are missing in the two samples. Unfortunately, we are not in this situation, since the ME-MACS mass limit is below the \Planck\ one and the mass measurements of ME-MACS are derived from a biased quantity such as the luminosity. To firmly test the effect of CC-bias we thus need to make use of numerical simulations (Sec. \ref{sec:sim}).

\subsection{Simulations}
\label{sec:sim}

\begin{figure}
\centering
\includegraphics[width=0.5\textwidth]{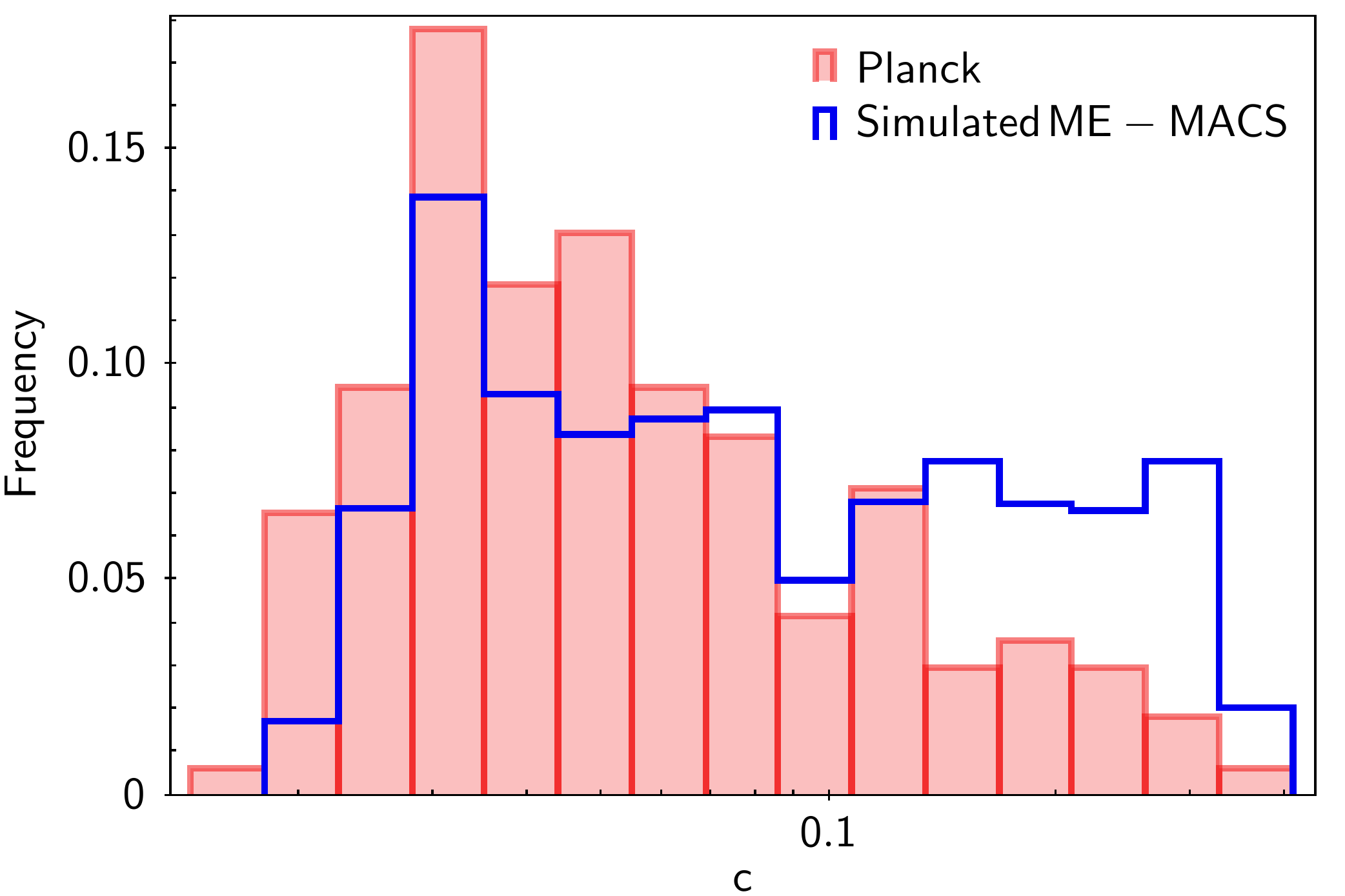}
\caption{Distribution of the concentration parameter in the \Planck\ sample (pink), used as an input in the simulation, and the output distribution of concentration parameters of the detected clusters in the ME-MACS simulation (blue). The vertical dashed line marks the threshold to separate CC ($c>0.075$) from NCC ($c<0.075$).}
\label{fig:simulcsb}
\end{figure}

\begin{figure}
\centering
\includegraphics[width=0.45\textwidth]{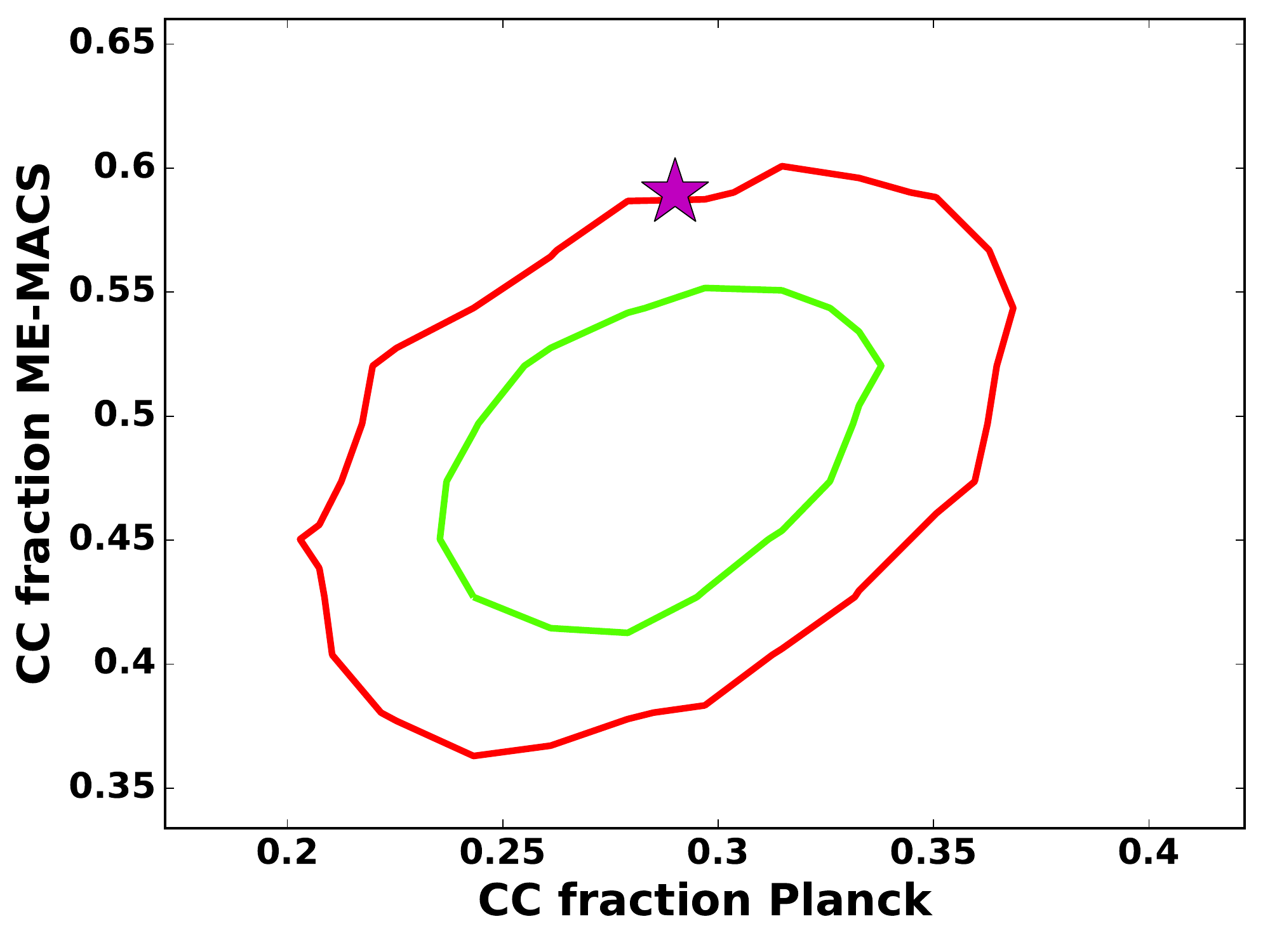}
\caption{Probability contours of the CC fraction drawn from 10,000 simulated populations of massive clusters, applying the \emph{Planck} and ME-MACS selection functions to the simulated data. The contours represent a containment of 68.3\% (red) and 99.7\% (green) of the simulations. The magenta star shows the true \emph{Planck} and ME-MACS CC fractions.}
\label{fig:ccfprob}
\end{figure}

We performed a dedicated simulation following an approach similar to \citet{eckert11} and tailored to reproduce the \Planck\ and ME-MACS selection criteria. 
The main idea of the procedure is to simulate a realistic population of clusters in the mass-redshift plane with a distribution of concentration parameters which follows the one in the \Planck\ sample,
apply the ME-MACS selection function to the simulated systems and measure the CC fraction in the ``detected'' simulated sample. We refer to \citet{eckert11} for the details of the simulation, here we recall the main steps and discuss the differences with respect to the previous approach. \\
We start by simulating a population of halos in the appropriate mass and redshift range. As opposed to \citet{eckert11}, who started from an X-ray luminosity function tailored for their sample, here we randomly draw halos according to the mass function of \citet{tinker08}. We then use the relation between core-excised X-ray luminosity and halo mass of \citet{mantz10} to calculate the expected luminosity of each halo. \\
To convert the core-excised luminosity into an integrated luminosity and overall flux, we associate a surface brightness profile to each simulated cluster. We underline that this is an improvement with respect to \citet{eckert11}, because we take into account that at a fixed mass, CC clusters are actually more luminous than NCC clusters (see discussion in \citealt{hudson10}). 
 In the original simulation, \citet{eckert11} used a fixed surface brightness template (a beta model for NCC and a double beta model for CC) and randomly chose between the two according to a fixed input CC fraction. Here we assume that the distribution of the concentration parameters of the \Planck\ sample is representative of the true distribution and we use the full measured distribution to draw a realistic distribution of surface brightness profiles. We assume that the surface brightness profile of each cluster can be approximated by a double beta model:
 \begin{equation}
 S(r)=S_1\left(\left[1+\left(\frac{r}{r_{c1}}\right)^2\right]^{-3\beta+1/2}+R\left[1+\left(\frac{r}{r_{c2}}\right)^2\right]^{-3\beta+1/2}\right),
 \label{eq_doublebeta}
 \end{equation} 
  where the ratio between the two beta models, $R$, is randomly selected from a list of values that reproduce the $c$ distribution of the \Planck\ sample, while $\beta$, $r_{c1}$ and $r_{c2}$ are fixed to the values which best represent the observed values in the \Planck\ sample ($\beta=0.64$, $r1=300$ kpc, $r2=30$ kpc) and $S_1$ is the overall normalization, which is set on-the-fly to reproduce the core-excised luminosity of each simulated halo. This approach allows us to simulate a population spanning a whole range of surface-brightness profiles, but the results do not change significantly if a fixed template  is used. After having selected a surface brightness profile, we can reproduce the integrated luminosity and the flux in the $0.1-2.4$ keV energy range for each simulated cluster, as described in \citet{eckert11}. A hidden assumption in this procedure is that the \Planck\ distribution of concentration parameter is representative of the cluster population at all redshifts, i.e. that the CC fraction does not evolve with time, in contrast with the recent result by \citet{mcdonald}. However, this evolution becomes strongly significant only at very high redshift ($z>0.7$), where we have very few objects both in our simulations and in the observed samples.  \\
 We then simulate the ME-MACS selection. As described in \citet{mann_ebe}, the starting point of the ME-MACS (and also MACS, \citealt{ebeling01}) is the RASS Bright Source Catalogue \citep{vogesBSC}, from which they selected candidate clusters with a flux-limit $f_{RASS}>10^{-12}\, \rm{ergs}\, \rm{cm}^{-2}\, \rm{s}^{-1}$. We can thus use the same procedure as in \citet{eckert11} to simulate the source detection as ME-MACS is a RASS-based flux-limited survey. We only apply the additional luminosity and redshift criteria in ME-MACS, namely $L_X>5\times10^{44}$ ergs s$^{-1}$ and $z>0.15$.  \\	
 We run our simulation with $10^7$ input halos, resulting in more than 15,000 detected clusters, for which we compute the concentration parameter. We compare the $c$ output distribution with the input one in Fig. \ref{fig:simulcsb}: it is apparent that a second peak of the distribution emerges at high concentration parameters (i.e. CC). While the starting population is described by the \Planck\ lognormal distribution, the output of the simulation is not described by a unimodal distribution anymore and a secondary peak emerges. Our simulation thus shows that the ``bimodality'' (i.e. presence of two peaks) of the cluster population between CC and NCC objects, which has been largely discussed in the literature (e.g. \citealt{cava09,pratt10}) is at least partly due to the CC-bias. \\ 
The CC fraction in the whole simulated sample is $48\%$, significantly larger than the fraction in the \Planck\ sample, but still lower than the measured value of the ME-MACS sample $(59 \pm 5\%)$. Since the \Planck\ and ME-MACS samples largely overlap and are drawn from the same underlying population, there is a strong covariance between the CC fraction measured in the two samples. This covariance needs to be taken into account to assess the significance of the difference between the two samples. To this aim, we perform another set of simulations in which the number of simulated halos reproduces the expected number of halos in the redshift and mass range of interest. We then apply at the same time the \Planck\ and ME-MACS selection functions to the simulated halos to draw realistic \Planck-like and ME-MACS-like cluster samples simultaneously and calculate the CC fraction in both. To implement the \Planck\ selection function, which is given as a function of the total SZ flux ($Y$) and the apparent opening angle \citep{PSZ2}, we use the $Y-M$ relation from \citet{cosmoPSZ1}.
We then repeat this procedure 10,000 times and compare the resulting CC bias values with the observed one. In Fig. \ref{fig:ccfprob} we show the 68.3\% and 99.7\% containment contours of the output values for the CC fraction. The figure clearly shows the strong covariance between the two measurements, which results from the fact that the two samples are not independent. In only 0.2\% of the cases we are finding that the two CC fractions are consistent with the observed ones simultaneously. \\
To summarize, our simulation reproduces qualitatively the properties of the ME-MACS sample and the presence of two peaks. It shows that the CC-bias certainly plays a large role in the difference between the \Planck\ and ME-MACS distribution, but at the same time suggests that is unlikely that CC-bias alone can account for the full difference.  However we should remind that our attempt to reproduce the effect of the CC bias, although sophisticated, is based on several assumptions and, as any simulation, cannot fully reproduce the complexity of the clusters population and of the X-ray and SZ selection. 
Moreover, as our samples are not fully complete, there is still the possibility that the difference may be fully explained by the CC bias, if we assume that the 12 missing clusters in the \Planck\ $z>0.15$ sample (Sec. \ref{sec:plancksample}) are CC (rising the CC fraction to 36\%) and all unobserved objects in ME-MACS (25, Sec. \ref{sec:macssample})  are NCC (lowering it to 47\%). 
Although we consider this hypothesis unlikely, we cannot exclude it given the incompleteness of our sample.

\section{Beyond the CC bias}
\label{sec:discussion}
In Sec. \ref{sec:ccbias}, we focused our attention on the role of CC-bias in explaining the difference between the \Planck\ and ME-MACS distribution. However, there are other mechanisms, both in the SZ and in the X-ray selection, that can contribute to the difference and that can be highlighted by the comparison of our samples. In this Section, we will discuss the role of possible biases against CCs in the \Planck\ survey (Sec. \ref{sec:antiCC}) and the effect due to a population of X-ray underluminous objects (Sec. \ref{sec:underL}) in the \Planck\ catalogue \citep{PSZ2}.

\subsection{An anti-CC bias in the \Planck\ SZ survey}
\label{sec:antiCC}
It has been suggested that the presence of bright radio galaxies at the centres of CC clusters may induce a bias against CCs in SZ surveys, since radio-sources could potentially influence the cluster detection and measurement of the SZ signal  (e.g. \citealt{sayers13_radiogal, lin15}) and are thus usually masked out in the SZ analysis. As discussed in \citet{PSZ1}, this bias is expected to be small in the \Planck\ survey, as the cluster detection is performed with HFI at high frequencies, where the emission of steep-spectrum radio sources is negligible with respect to the SZ effect. Nonetheless, we tried to estimate this possible  bias which could in principle contribute to the residual difference that we found in Sec. \ref{sec:sim}.
The point source mask used in the \Planck\ analysis was built starting from the \Planck\ compact source catalogues (PCCS, \citealt{PCCS}) at several frequencies, excluding a circle of radius $2.13$ FWHM around point sources detected with signal-to-noise ratio larger than ten \citep{PSZ1}. Starting from the Meta-Catalogue of X-ray detected Clusters (MCXC, \citealt{mcxc}), we looked for known clusters whose position is within a radius of $2.13$ FWHM around bright point sources in at least one out of the six HFI frequencies, finding 57 candidate ``missing'' clusters. However, most of these objects have low masses and should not be detectable by \Planck. Only for 6 objects (namely Perseus, Cygnus A, Abell 780, RXC J$1130.3-1434$, RXC J$1025.9+1241$, Abell S1111) the masses in the MCXC catalogue are comparable with the masses of clusters in the PSZ1-cosmo sample at the same redshift and they could thus be detected in the survey if they were not behind the mask. Perseus, Cygnus-A and Abell 780 are well known CC clusters, while we could not find any literature information about the core state of the remaining three objects,  whose expected mass is furthermore close to the limit of the selection function in the mass-redshift plane and may thus be not detected by \Planck, also for statistical reasons.   
We can thus roughly estimate that the \Planck\ catalogue is missing 3-6 objects because of radio sources, and assuming that they are all CC,  the corrected CC fraction would be $30-31\%$. The bias due to radio-galaxies in CC is thus only $1-2\%$, smaller than the statistical error on the CC fraction in the \Planck\ sample, and  not sufficient to reach the CC fraction of 38\% that would be needed to reproduce the ME-MACS fraction with our simulation of the CC bias.  \\
Another possible reason for which the \Planck\ SZ survey may be biased against CC is that through the SZ effect we may in principle detect more easily disturbed merging clusters, where the SZ signal may be enhanced by shock fronts propagating in the ICM. Indeed, \citet{sommerbasu} showed that the SZ signal within $R_{500}$ in simulated clusters is boosted after a merger on a time-scale of a few Gyr.
The selection function and its dependence on the dynamical state of the \Planck\ SZ survey has been tested in \citet{PSZ2} with Monte-Carlo simulations, by injecting simulated clusters with different $y$ maps in the \Planck\ sky maps and running the detection algorithms, showing that the cluster morphology does not impact significantly the source detection. This result is not unexpected if we take into account the large beams of the \Planck\ frequency channels: similarly to what happens to the peaked pressure profile of CC clusters, over-pressurized regions due to shocks are smoothed out by the \Planck\ moderate spatial resolution. Moreover, \Planck\ is more sensitive to the behaviour of the pressure profiles at large scales than to the smaller-scale physics (such as cool cores or shock) and it measures the SZ signal on scales larger than $R_{500}$ (i.e. the region studied by \citealt{sommerbasu}). \\
 Recently, \citet{nurgaliev16} suggested that \Planck\ may be more sensitive to pairs or triplets of galaxy clusters, because of its large beam capturing an inflated signal from multiple objects and therefore may be biased towards merging systems. While it is certainly true that \Planck\ has detected a few of these objects that received a lot of attention in the literature \citep{planck_early_IX, pip_VI}, in the high purity PSZ1 cosmology sample that we analyzed in this paper and in Paper I, we do not have a large number of these objects. 
 Moreover, clusters in multiple systems are not necessarily disturbed NCC objects: for instance, the brightest member in the \Planck\ discovered supercluster PLCK G$214.6+37.0$ features a prominent surface brightness peak associated to the BCG \citep{pip_VI}, which would led us to classify it as a CC relaxed object.\\
 Last but not least, if there were a systematic difference between the pressure profile of CC and NCC clusters at $R \gtrsim R_{500}$, with NCC clusters showing flatter profile than CC clusters similarly to what observed in the gas density distribution \citep{eckert12}, NCC could possibly have a larger SZ signal at large scales making them easier to detect in SZ. However, the analysis of the pressure profiles of samples of galaxy clusters both with \Planck\ \citep{pip_V} and with Bolocam \citep{sayers13_Pprof} show only a moderate difference at large scales and with a large dispersion. Indeed, if we assume the best fit models for CC and NCC objects in the analysis of \Planck\ pressure profiles \citep{pip_V} and we integrate them to measure the SZ signal at $5R_{500}$, the derived values differ only by 2\%. Nonetheless, the SZ flux at $5R_{500}$ depends strongly on the shape of the pressure profile: if we assume a combination of parameters consistent at 68\% with the best fit model but with a flatter outer slope $\beta=3.2$ (basing on Fig. 5 in \citealt{pip_V}), the derived $Y_{5R500}$ would be 12\% larger than the value with the mean CC profile. We underline that the sample of clusters for which the \Planck\ pressure profile has been measured is not SZ-selected, as it is composed of early \Planck\ detections already known in X-rays and with available \xmm\ data \citep{pip_V}, thus the derived pressure profile may not be representative of the cluster population. While present data do not allow to provide support to the hypothesis of an anti-CC bias in \Planck\, more detailed studies on larger and well-defined samples are needed to reduce the uncertainties and to firmly assess the shape of the CC and NCC pressure profiles and their role on the SZ detection procedures.\\
 
\subsection{X-ray underluminous clusters}
\label{sec:underL}
\begin{figure}
\centering
\includegraphics[width=0.5\textwidth]{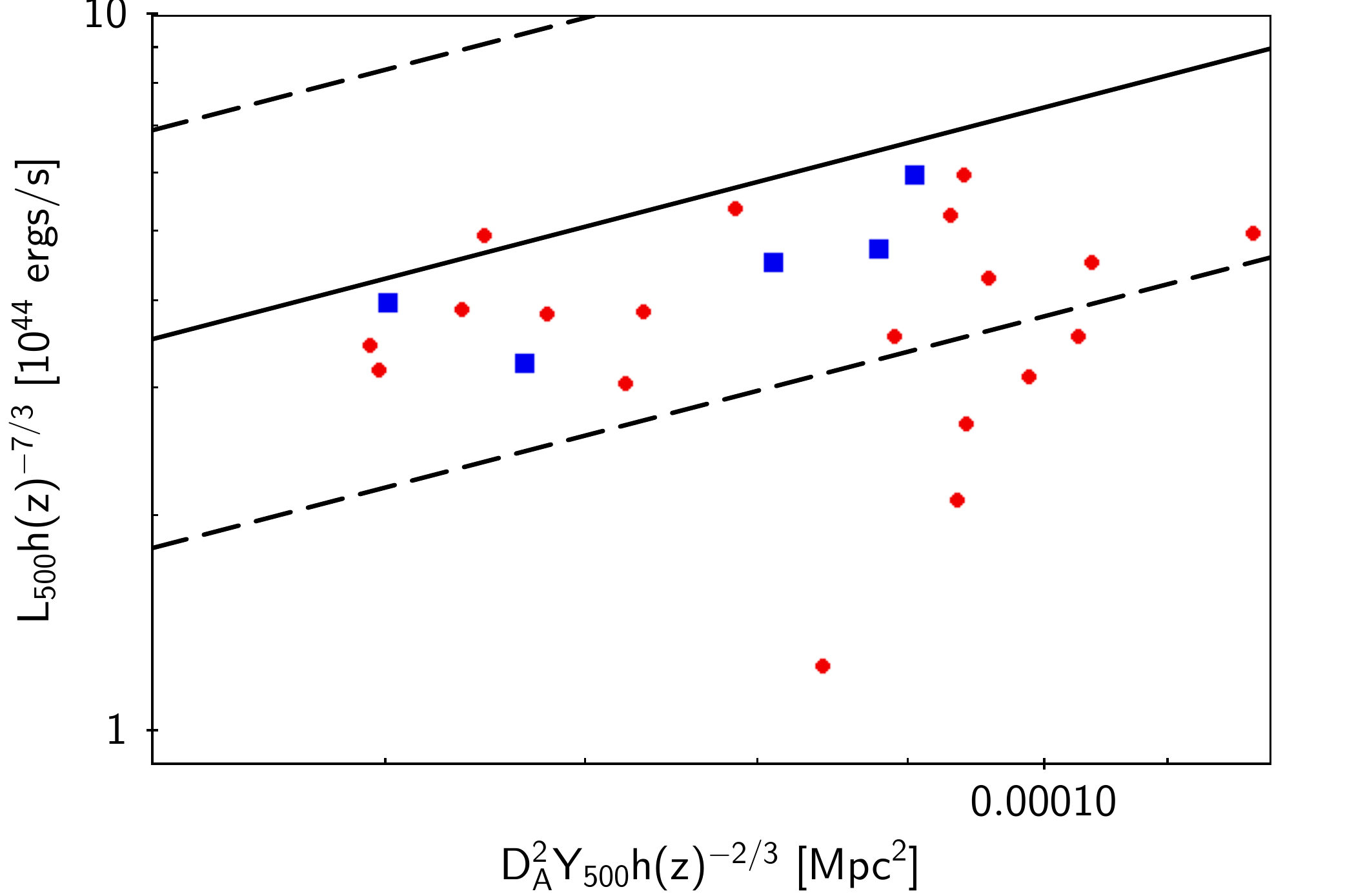}
\caption{Distribution in the $Y_{500}-L_{500}$ plane of the 24 missing ME-MACS clusters, compared with the best-fit scaling relation (black continuous line) and its dispersion ($\pm 2\sigma$, dashed lines) estimated in \citet{PSZ2}.. We mark with blue squares CC clusters and with red points NCC.}
\label{fig:underluminous}
\end{figure}

One unexpected result of the \Planck\ SZ survey has been the discovery of a population of X-ray under-luminous clusters \citep{PSZ2}. These systems feature an X-ray luminosity which is well below the value that could be expected through scaling relations from the SZ signal, while their optical richness is in agreement with expectations. This population was highlighted at low redshift and in the SDSS sky area, but it possibly extends also to other redshift ranges and sky regions.  If this population exists also in our \Planck\ sample and if all, or most, under-luminous clusters are classified as NCC, it could contribute in explaining the difference between the CC fraction in the \Planck\ sample and in ME-MACS. \textbf{As these objects by definition obey to a different $L-M$ scaling relation than the one we used in Sec. \ref{sec:sim}, their presence is not accounted for in our simulation.} \\
One method to highlight this population in our \Planck\ sample is to look at the clusters which should have been detected also in ME-MACS but are not (complementary to what we showed in \ref{sec:notplanck}). We thus select all clusters in the \Planck\ sample which lie in the sky region covered by ME-MACS, have $z>0.15$ and an expected luminosity larger than the ME-MACS threshold (see \citealt{mann_ebe} for details) but are missing in the ME-MACS sample. To estimate the expected luminosity, we convert the SZ signal $Y_{500}$ in the \Planck\ catalogue into $L_{500}$, using the $L_{500}-Y_{500}$ relation obtained in \citet{PSZ2}. 
 With this method, we find 24 missing clusters in ME-MACS, most of which (19) are NCC objects.
 In Fig. \ref{fig:underluminous}, we compare their measured luminosity\footnote{For most objects, we used the luminosity in the MCXC catalogue \citep{mcxc}, while for 5 objects we measured the luminosity directly from the \chandra\ data.} $L_{500}$ as a function of their SZ signal, with the scaling relation and its scatter, calibrated on \Planck\ clusters by \citet{PSZ2}. We notice that almost all objects lie below the expected relation and some of them below twice the intrinsic scatter, which would lead to their classification as ``underluminous'' objects, following \citet{PSZ2}. 
According to the concentration parameter, all the most deviant objects are classified as NCC. We noticed that in a few cases the measured luminosity is above the selection threshold of the ME-MACS sample ($5\times 10^{44}\ \rm{ergs}\ \rm{s}^{-1}$). However, we used luminosities within $R_{500}$, while the luminosity used in the selection of the ME-MACS sample is estimated in the RASS detection cell. Indeed, one of the most luminous clusters in Fig. \ref{fig:underluminous} is A115N, which has $L_{500}=7.5\times 10^{44} \ \rm{ergs}\ \rm{s}^{-1}$ in MCXC, but with $L_{RASS,det}=4.4 \times 10^{44} \ \rm{ergs}\ \rm{s}^{-1}$ it fails to make the luminosity cut in ME-MACS (H. Ebeling, private communication). \\
The population of X-ray under-luminous clusters is thus likely present also in the \Planck\  sample we are analyzing. It is intriguing that candidate X-ray underluminous clusters in our sample are almost all classified ad NCC: if this population, which is missing in X-ray surveys but is detected in SZ,  is composed of disturbed NCC clusters, they could certainly contribute to the residual difference between the \Planck\ and ME-MACS distribution of concentration parameters. At the moment, little is known about these objects, and it is unclear if they are truly X-ray underluminous for their mass or if their SZ signal is artificially boosted. New observations, both in X-rays and possibly in SZ, are needed to assess the origin of this class of objects. A systematic analysis of their properties and the cool-core state of X-ray under-luminous clusters is beyond the scope of this paper and will be presented in a forthcoming work with new dedicated data (Rossetti et al. in prep.).

\section{Summary and conclusions}
In this paper, we studied the cool core state of a SZ-selected sample of galaxy clusters, the cosmological sample of the first \Planck\ SZ catalogue \citep{PSZ1}, using as indicator the concentration parameter \citep{santos08}. Our results are summarized as follows.
\begin{itemize}
\item The distribution of the concentration parameters in the \Planck\ sample features a single peak at low values of $c$. The fraction of CC clusters ($c > 0.075$) is $(29 \pm 4) \%$.
\item We do not find indications of evolution of the CC fraction by dividing our sample in two redshift bins. Our result does not contradict previous detections which report evolution in a redshift range ($z>0.3$) which is poorly sampled by our catalogue \citep{mcdonald}. We find an indication of a larger CC fraction in higher mass systems, as reported also in \citet{mantz15}, but only at low significance ($1.5 \sigma$).
\item We compared the distribution of the concentration parameter with the one of the X-ray selected ME-MACS sample \citep{mann_ebe}. The distributions are significantly different with a $1.7\times 10^{-7}$  probability that they are drawn from the same population of objects. Indeed, ME-MACS hosts a much larger fraction of CC objects: $(59 \pm 5) \%$. 
\item The distributions of concentration parameters in ME-MACS shows two peaks and is well described by two gaussians. This double peaked distribution, which is observed also in other X-ray selected samples and with other cool-core indicator \citep{cava09,pratt10} has opened a debate in the literature whereas the cluster population is bimodal or not. However, our \Planck\ sample is better described by a single lognormal distribution. We showed with simulations that a secondary peak at high concentration parameters emerges in X-ray flux-limited samples as a consequence of the CC-bias and the presence of two peaks may thus not be an intrinsic property of the cluster population.  
\item Among the X-ray selected samples available in the literature, ME-MACS is the one with the mass and redshift distributions more similar to the \Planck\ sample \citep{rossetti16}. Nonetheless, we compared the $c$ distribution in \Planck\ with the one of other X-ray samples and found them to be significantly different, having CC fractions in the range $56-74 \%$. We also compared our distribution with the one in the SZ selected sample of SPT clusters \citep{mcdonald}, finding them to be consistent with a comparable CC fraction ($29 \pm 7$ \%) in the common redshift range.
\item A possible origin of the discrepancy between the CC fraction in SZ-selected and X-ray selected samples is the CC-bias \citep{eckert11}. We tested this hypothesis with simulations of the CC bias: starting from a realistic population of clusters with the distribution of concentration parameters in the \Planck\ sample, we simulate the ME-MACS selection and measure the CC fraction in the output sample (Sec. \ref{sec:sim}). Starting from a CC-fraction of 29\% in the input population, we obtain a CC fraction of 48\% in the output sample,  showing that CC-bias plays a large role in the difference between the two samples. Nonetheless,  according to our simulation, the probability of obtaining simultaneously two CC-fractions of 29\% in \Planck\ and 59\% in ME-MACS is only $0.2\%$.
\item We considered several mechanisms that could also possibly affect SZ surveys to be biased against CC, namely the presence of radio galaxies in CCs, the role of shocks in increasing the SZ signal, the large \Planck\ beam favoring the detection of multiple disturbed objects and a difference in the pressure profile at large radii. However, none of them seem sufficient to explain the difference between the observed CC-fraction in ME-MACS and the one in \Planck.
\item We noticed that the \Planck\ sample host a population of objects, which according to their expected luminosity (from $L-Y$ scaling relation) should be present also in ME-MACS, but are not since their observed luminosity is below the luminosity cut in that sample. Most of these X-ray underluminous objects are classified as NCC. The presence of this population of clusters, whose origin and properties are still unclear, in the \Planck\ sample could possibly contribute to the  difference.
\end{itemize}

\section*{Acknowledgements}

We dedicate this paper to the memory of our colleague Y.Y. Zhang: we are indebted to her work \citet{zhang11} for the analysis discussed in Paper I, on which the present paper is based.
We thank H. Ebeling for useful information about the ME-MACS sample and T. Reiprich for providing the concentration parameters of HIFLUGCS clusters. 




\bibliographystyle{mnras}
\bibliography{mybib}




\begin{appendix}

\section{Correlation with $D_{X-BCG}$}
\label{app:correlation}
\begin{figure}
\centering
\includegraphics[width=0.5\textwidth]{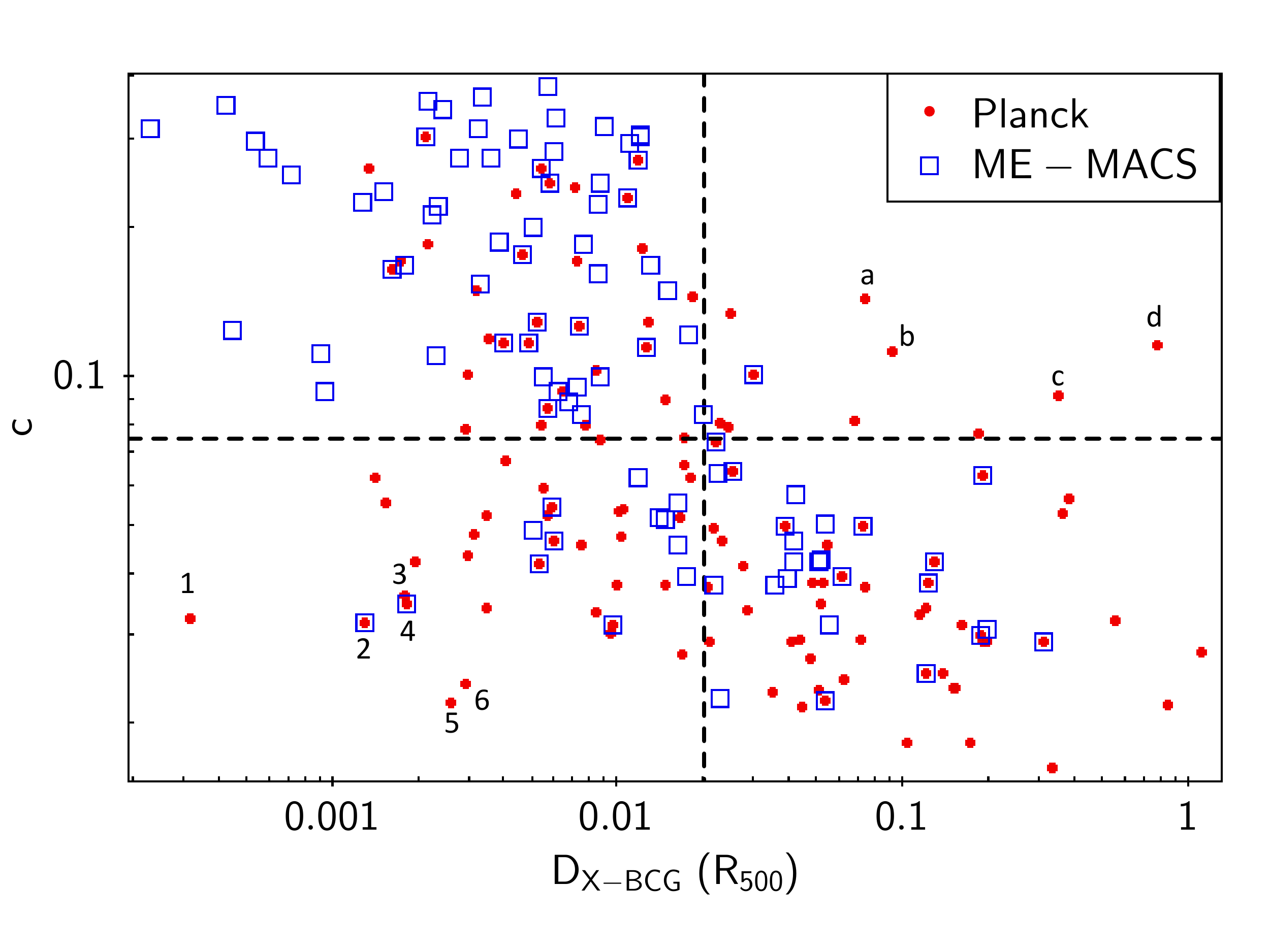}
\caption{Distribution in the $D_{X-BCG}-c$ plane of \Planck\ (red circles) and ME-MACS (blue open squares) clusters. Dashed lines mark the separation between CC/NCC and relaxed/disturbed clusters.  Objects in the upper left quadrant are relaxed CC while in the lower quadrant lie disturbed NCC systems.}
\label{fig:correlation}
\end{figure}

As discussed in Sec. \ref{sec:intro}, the concentration parameter is an indicator of the presence of a CC, while the indicator used in Paper I, i.e. $D_{X-BCG}$ the projected distance between the X-ray peak and the BCG, is an indicator of dynamical activity. CC are usually found in dynamically relaxed systems and $D_{X-BCG}$ has been shown to correlate with thermodynamical indicators of the cool core state \citep{sanderson09}. Here, we test the correlation between $D_{X-BCG}$ and $c$ for the \Planck\ and ME-MACS sample. For \Planck\ we used the values in Paper I for the 122 common clusters, while for ME-MACS we used the values provided for the full sample in \citet{mann_ebe}. The correlation plot is shown in Fig. \ref{fig:correlation}, where we also show the threshold values that we used to classify clusters in CC/NCC here  and relaxed/disturbed in Paper I ($D_{X-BCG}=0.02R_{500}$). 
We performed the Spearman and Kendall correlations test on both samples separately and on the joint sample. The results are shown in Table \ref{tab:correlation}. In both datasets separately and in the joint one we find a significant anti correlation between the two indicators, with most relaxed clusters being also CC and disturbed ones being NCC. The correlation is stronger for ME-MACS than for \Planck\, which hosts a larger number of outliers, i.e. clusters classified as relaxed but without a CC (possibly for projection effects) and disturbed objects with a CC. 
We investigated one by one the most outstanding outliers in the plot, that we label with numbers and letters in Fig. \ref{fig:correlation}, with the aim of trying to understand if their presence in \Planck\ but not in ME-MACS may be related to selection effects. The lower left quadrant of the plot contains clusters classified as ``relaxed NCC'': as discussed in Paper I, we expect that $7.5\%$ of the clusters classified ad relaxed by $D_{X,BCG}$ are in fact disturbed object where the separation between the X-ray peak and the BCG occurs mainly along the line of sight. Moreover an intrinsic limitation of the dynamical indicator $D_{X-BCG}$ is that not all mergers, and not all phases of the mergers, cause an offset between the BCG and the X-ray peak. The most deviating outliers in this panel are: (1) A2147 (\citealt{hudson10}, $z=0.03$), (2) A1758N ( \citealt{david04}, $z=0.27$, also in ME-MACS), (3) A3266 ( \citealt{finoguenov06}, $z=0.05$), (4) A697 ( \citealt{girardi06}, $z=0.28$, also in ME-MACS), (5) A119 (\citealt{hudson10}, $z=0.05$), and (6) A1437 ($z=0.13$, little X-ray information is available in the literature, our own analysis shows a disturbed and elongated morphology). Visual inspection of their X-ray images shows that they are all clearly disturbed objects undergoing mergers, as also supported by the literature. Most of them are simply not in ME-MACS because they are local systems ($z<0.15$), while the only two objects at $z>0.15$ are also found in ME-MACS. \\
The upper right panels contains object classified as "disturbed CC" and is populated mainly by \Planck\ objects. The most deviating objects are: (a) RXC J0232.2-4420 (see image in the ACCEPT archive \citealt{cava09}, $z=0.28$), (b) RXC J0638.7-5358  (see image in the ACCEPT archive \citealt{cava09}, $z=0.22$), (c) SPT-CL J0411-4819 (\citealt{mcdonald}, $z=0.43$) and (d) ACT-CL J0102-4915 a.k.a. El Gordo (\citealt{menanteau_elgordo}, $z=0.89$). It is interesting to note that all these clusters feature a significant surface brightness peak in an overall disturbed X-ray morphology. El Gordo is probably the most striking example: \citet{menanteau_elgordo} show that it is undergoing a major merger but it preserves a bright region with cool, low entropy and high metal abundance gas, likely the ``cool core remnant'' \citep{rossetti10} of one of the merging substructures. Similar systems should be in principle easily detected also in X-ray surveys: however they are not in ME-MACS simply because they do not fall in the sky area surveyed by MACS ($\delta > -40$, \citealt{mann_ebe}). \\
Analysis of the outliers in this relation could have been in principle very useful to suggest possible selection effects but  we are unfortunately limited by the incomplete spatial and redshift overlap of the two surveys, as discussed also in Sec. \ref{sec:notplanck}.

\begin{table}
 \centering
\begin{tabular}{l | c  c |  c c }
\hline
Sample & Spearman $\rho$ &  $p_0$ & Kendall $\tau$ &$p_0$ \\
  \hline
 \Planck\ & $-0.43$  & $6 \ 10^{-7}$ & $-0.30$ & $8 \ 10^{-7}$  \\
 ME-MACS & $-0.74$ & $2\ 10^{-19}$ & $-0.54$ & $4 \ 10^{-16}$\\
 Joint & $-0.60$ & $2\ 10^{-23}$ & $-0.43$ & - \\
\hline
 \end{tabular}
\caption{Output of correlation tests}
\label{tab:correlation}
\end{table}

\end{appendix}

\clearpage
\onecolumn

\landscape
\begin{footnotesize}
\setcounter{table}{0}
\begin{longtable}{c c c c c c c c c c}

\hline
 INDEX & NAME & Alt. Name & R.A.$_X$ & Dec.$_X$ & Redshift & $M_{500}$ & $c$ & $\sigma_c$ & Obs. ID \\
   &   &   &   &    &   & $10^{14} M_\odot$ &  &   &  \\
\hline
\endhead

\hline
\endfoot

\hline
\caption{Properties of the clusters in our \Planck\ sample. Col. [1] is the INDEX in the PSZ1 catalogue, col. [2] the \Planck\ name, col. [3] provides an alternative name, Cols. [4] and [5] are the coordinates of the X-ray peak that we used to measure the concentration parameter, col. [6] and col[7] are the redshift and mass as provided in the \Planck\ catalogue, col. [8] and col. [9] are the concentration parameter  and its error, while in col [10] we list the ID of the observations  used in our analysis (those starting with ``0'' are \xmm\ data). }
\endlastfoot

6 & PSZ1 G002.77-56.16 & RXC J2218.6-3853 & 22:18:39.66 & -38:53:59.0 & 0.141 & 4.4 & 0.0585 & 0.0032 & 15101\\
  10 & PSZ1 G003.93-59.42 & RXC J2234.5-3744 & 22:34:27.53 & -37:43:57.0 & 0.150 & 6.6 & 0.0332 & 0.0016 & 15303\\
  17 & PSZ1 G006.45+50.56 & RXC J1510.9+0543 & 15:10:56.09 & +05:44:40.8 & 0.076 & 6.8 & 0.1701 & 0.0019 & 6101\\
  18 & PSZ1 G006.68-35.52 & RXC J2034.7-3548 & 20:34:48.74 & -35:50:54.6 & 0.089 & 4.0 & 0.0216 & 0.0017 & 12274\\
  23 & PSZ1 G008.33-64.74 & ACO S 1077 & 22:58:48.32 & -34:47:59.1 & 0.312 & 7.7 & 0.0478 & 0.0019 & 1562\\
  24 & PSZ1 G008.42-56.34 & RXC J2217.7-3543 & 22:17:45.55 & -35:43:22.4 & 0.148 & 4.8 & 0.0927 & 0.0050 & 15116\\
  26 & PSZ1 G009.02-81.22 & RXC J0014.3-3023 & 00:14:19.04 & -30:23:30.0 & 0.306 & 9.5 & 0.0223 & 0.0009 & 2212,7915,8477,8557\\
  54 & PSZ1 G021.10+33.24 & RXC J1632.7+0534 & 16:32:46.94 & +05:34:32.1 & 0.151 & 7.9 & 0.3035 & 0.0013 & 499,6104,7940\\
  74 & PSZ1 G028.77-33.56 & RXC J2048.1-1750 & 20:48:10.80 & -17:51:21.6 & 0.147 & 4.7 & 0.0189 & 0.0011 & 0654440401\\
  76 & PSZ1 G029.10+44.54 & RXC J1602.3+1601 & 16:02:16.32 & +15:58:12.0 & 0.035 & 2.9 & 0.0325 & 0.0007 & 0505210601\\
  92 & PSZ1 G033.43-48.44 & RXC J2152.4-1933 & 21:52:21.26 & -19:32:52.1 & 0.094 & 4.1 & 0.1446 & 0.0028 & 4202\\
  93 & PSZ1 G033.84+77.17 & RXC J1348.8+2635 & 13:48:52.54 & +26:35:32.0 & 0.062 & 4.5 & 0.1810 & 0.0007 & 10898,10899,12026,18424,5289,5290\\
  94 & PSZ1 G034.03-76.59 & RXC J2351.6-2605 & 23:51:39.36 & -26:05:01.9 & 0.226 & 6.8 & 0.1642 & 0.0041 & 2214\\
  108 & PSZ1 G039.81-39.96 & RXC J2127.1-1209 & 21:27:09.13 & -12:10:03.0 & 0.175 & 5.7 & 0.0183 & 0.0018 & 15103,16292\\
  113 & PSZ1 G040.63+77.13 & RXC J1349.3+2806 & 13:49:23.84 & +28:06:32.1 & 0.074 & 3.2 & 0.0753 & 0.0046 & 15153\\
  120 & PSZ1 G042.85+56.63 & RXC J1522.4+2742 & 15:22:29.42  & +27:42:20.23  & 0.072 & 4.3 & 0.0794 & 0.0010 & 3182\\
  122 & PSZ1 G044.24+48.66 & RXC J1558.3+2713 & 15:58:20.88 & +27:13:44.2 & 0.089 & 8.8 & 0.0802 & 0.0003 & 1196,1228,15186,16564,16565,5005\\
  125 & PSZ1 G044.77-51.30 & RXC J2214.9-1400 & 22:14:57.28 & -14:00:12.9 & 0.502 & 8.4 & 0.0666 & 0.0043 & 3259,5011\\
  137 & PSZ1 G046.09+27.16 & RXC J1731.6+2251 & 17:31:39.64 & +22:52:11.1 & 0.388 & 7.9 & 0.0187 & 0.0034 & 3281\\
  140 & PSZ1 G046.48-49.42 & RXC J2210.3-1210 & 22:10:18.98 & -12:09:50.5 & 0.084 & 4.4 & 0.0341 & 0.0020 & 8271\\
  141 & PSZ1 G046.90+56.48 & RXC J1524.1+2955 & 15:24:07.53 & +29:53:16.8 & 0.114 & 5.4 & 0.0217 & 0.0011 & 4965\\
  153 & PSZ1 G049.22+30.84 & RXC J1720.1+2637 & 17:20:10.52 & +26:37:47.0 & 0.164 & 6.3 & 0.1258 & 0.0022 & 4361\\
  164 & PSZ1 G053.42-36.25 & RXC J2135.2-0102 & 21:35:11.05 & -01:02:53.2 & 0.330 & 7.5 & 0.0386 & 0.0030 & 11710,16285\\
  166 & PSZ1 G053.52+59.52 & RXC J1510.1+3330 & 15:10:13.36 & +33:30:39.5 & 0.112 & 4.9 & 0.0314 & 0.0004 & 12885,12886,13192,13193,2204\\
  174 & PSZ1 G055.58+31.87 & RXC J1722.4+3208 & 17:22:27.32 & +32:07:57.4 & 0.224 & 7.3 & 0.1167 & 0.0031 & 5007\\
  177 & PSZ1 G055.95-34.87 & RXC J2135.2+0125 & 21:35:18.29 & +01:25:27.8 & 0.231 & 6.9 & 0.0296 & 0.0035 & 15097\\
  180 & PSZ1 G056.79+36.30 & RXC J1702.7+3403 & 17:02:42.72 & +34:03:40.4 & 0.095 & 4.0 & 0.1010 & 0.0009 & 4179\\
  181 & PSZ1 G056.94-55.06 & RXC J2243.3-0935 & 22:43:21.19 & -09:35:37.2 & 0.446 & 10. & 0.0252 & 0.0033 & 3260\\
  183 & PSZ1 G057.28-45.37 & RXC J2211.7-0349 & 22:11:45.87 & -03:49:47.3 & 0.397 & 9.2 & 0.1006 & 0.0042 & 3284\\
  185 & PSZ1 G057.63+34.92 & RXC J1709.8+3426 & 17:09:49.15 & +34:27:11.3 & 0.080 & 3.6 & 0.0233 & 0.0021 & 12284\\
  224 & PSZ1 G067.19+67.44 & RXC J1426.0+3749 & 14:26:03.12 & +37:49:24.9 & 0.171 & 6.9 & 0.0632 & 0.0014 & 3593,542\\
  238 & PSZ1 G071.63+29.78 & RXC J1747.2+4512 & 17:47:08.99 & +45:12:44.9 & 0.156 & 4.3 & 0.0207 & 0.0028 & 15118\\
  242 & PSZ1 G072.61+41.47 & RXC J1640.3+4642 & 16:40:19.94 & +46:42:45.3 & 0.228 & 11. & 0.0316 & 0.0008 & 896\\
  248 & PSZ1 G073.98-27.83 & RXC J2153.5+1741 & 21:53:36.81 & +17:41:43.1 & 0.232 & 9.4 & 0.1287 & 0.0011 & 4193\\
  252 & PSZ1 G075.71+13.51 & RXC J1921.1+4357 & 19:21:10.90 & +43:56:45.4 & 0.055 & 8.5 & 0.0466 & 0.0003 & 15187,3231\\
  256 & PSZ1 G077.89-26.62 & RXC J2200.8+2058 & 22:00:52.51 & +20:58:04.9 & 0.146 & 5.4 & 0.0677 & 0.0033 & 3247\\
  268 & PSZ1 G081.01-50.92 & RXC J2311.5+0338 & 23:11:33.25 & +03:38:08.2 & 0.299 & 7.5 & 0.0644 & 0.0032 & 11730,3288\\
  291 & PSZ1 G085.98+26.69 & RXC J1819.9+5710 & 18:19:54.00 & +57:09:21.4 & 0.179 & 4.2 & 0.0279 & 0.0036 & 15131,16579\\
  297 & PSZ1 G087.03-57.37 & RXC J2337.6+0016 & 23:37:37.92 & +00:16:03.3 & 0.277 & 6.9 & 0.0302 & 0.0024 & 11728,3248\\
  313 & PSZ1 G091.82+26.11 &  &18:31:08.59 & +62:14:12.96   & 0.822 & 7.4 & 0.0344 & 0.0067 & 18285\\
  319 & PSZ1 G092.67+73.44 & RXC J1335.3+4059 & 13:35:16.29 & +41:00:00.4 & 0.227 & 8.2 & 0.0384 & 0.0025 & 3591\\
  325 & PSZ1 G093.93+34.92 & RXC J1712.7+6403 & 17:12:39.96 & +64:03:16.8 & 0.080 & 5.1 & 0.0184 & 0.0005 & 894\\
  341 & PSZ1 G097.72+38.13 & RXC J1635.8+6612 & 16:35:51.25 & +66:12:36.5 & 0.170 & 6.4 & 0.0417 & 0.0014 & 1454,1666,553\\
  388 & PSZ1 G106.84-83.24 & RXC J0043.4-2037 & 00:43:24.23 & -20:37:33.7 & 0.292 & 9.1 & 0.0498 & 0.0034 & 9409\\
  389 & PSZ1 G107.14+65.29 & RXC J1332.7+5032 & 13:32:38.51 & +50:33:43.1 & 0.279 & 7.9 & 0.0318 & 0.0018 & 2213\\
  407 & PSZ1 G110.99+31.74 & RXC J1703.8+7838 & 17:03:00.60 & +78:38:59.4 & 0.058 & 6.3 & 0.0293 & 0.0003 & 1386,16129,16514,16515,16516\\
  411 & PSZ1 G112.48+57.02 & RXC J1336.1+5912 & 13:36:08.42 & +59:12:23.1 & 0.070 & 3.1 & 0.0517 & 0.0024 & 12282\\
  415 & PSZ1 G113.84+44.33 & RXC J1414.2+7115 & 14:13:54.32 & +71:17:40.0 & 0.224 & 5.0 & 0.0292 & 0.0041 & 15129\\
  417 & PSZ1 G114.29+64.91 & RXC J1315.1+5149 & 13:15:05.08 & +51:49:03.4 & 0.283 & 5.9 & 0.0486 & 0.0020 & 15123,16126\\
  419 & PSZ1 G114.78-33.72 & RXC J0020.6+2840 & 00:20:37.55 & +28:39:32.8 & 0.093 & 3.8 & 0.0492 & 0.0037 & 15164\\
  422 & PSZ1 G114.99+70.36 & RXC J1306.9+4633 & 13:06:49.79 & +46:33:29.8 & 0.225 & 6.1 & 0.0417 & 0.0034 & 11725,3244\\
  423 & PSZ1 G115.20-72.07 & RXC J0041.8-0918 & 00:41:50.26 & -09:18:11.3 & 0.055 & 4.9 & 0.1483 & 0.0004 & 15173,15174,16263,16264,904\\
  454 & PSZ1 G124.20-36.47 & RXC J0055.9+2622 & 00:55:50.42 & +26:24:35.9 & 0.197 & 7.2 & 0.1705 & 0.0014 & 13458,13459,15578,15581,3233\\
  459 & PSZ1 G125.68-64.12 & RXC J0056.3-0112 & 00:56:20.16 & -01:14:34.1 & 0.044 & 3.3 & 0.0220 & 0.0005 & 4180,7918\\
  460 & PSZ1 G125.72+53.87 & RXC J1236.9+6311 & 12:36:58.66 & +63:11:13.3 & 0.301 & 5.9 & 0.0565 & 0.0030 & 15127,7938\\
  482 & PSZ1 G134.73+48.89 & RXC J1133.2+6622 & 11:33:14.64 & +66:22:48.0 & 0.115 & 3.5 & 0.0663 & 0.0023 & 0083150401\\
  502 & PSZ1 G139.17+56.37 & RXC J1142.5+5832 & 11:42:23.70 & +58:31:53.8 & 0.321 & 7.1 & 0.0231 & 0.0026 & 15136\\
  503 & PSZ1 G139.61+24.20 &  & 06:21:48.95 & +74:42:04.8 & 0.266 & 7.0 & 0.1744 & 0.0050 & 15139,15297\\
  513 & PSZ1 G143.28+65.22 & RXC J1159.2+4947 & 11:59:14.83 & +49:47:33.2 & 0.350 & 7.3 & 0.0384 & 0.0045 & 15119\\
  530 & PSZ1 G149.21+54.17 & RXC J1058.4+5647 & 10:58:26.86 & +56:47:37.3 & 0.136 & 6.2 & 0.0302 & 0.0026 & 13376\\
  532 & PSZ1 G149.55-84.16 & RXC J0102.7-2152 & 01:02:41.72 & -21:52:53.9 & 0.056 & 3.0 & 0.2395 & 0.0012 & 13518,3183,3710,9897\\
  533 & PSZ1 G149.75+34.68 & RXC J0830.9+6551 & 08:30:58.87 & +65:50:17.4 & 0.181 & 8.2 & 0.0498 & 0.0019 & 3586\\
  535 & PSZ1 G150.56+58.32 & RXC J1115.2+5320 & 11:15:15.09 & +53:19:58.0 & 0.469 & 7.9 & 0.0318 & 0.0036 & 3253,5008,5350\\
  558 & PSZ1 G159.81-73.47 & RXC J0131.8-1336 & 01:31:52.76 & -13:36:41.4 & 0.206 & 8.1 & 0.0468 & 0.0025 & 3579,522\\
  560 & PSZ1 G161.39+26.24 & RXC J0721.3+5547 & 07:21:31.44 & +55:45:43.2 & 0.038 & 2.0 & 0.0678 & 0.0008 & 0504320101\\
  567 & PSZ1 G163.69+53.52 & RXC J1022.5+5006 & 10:22:28.25 & +50:06:22.3 & 0.158 & 4.9 & 0.0595 & 0.0037 & 15105\\
  572 & PSZ1 G165.06+54.13 & RXC J1023.6+4907 & 10:23:39.84 & +49:08:37.7 & 0.143 & 4.6 & 0.0458 & 0.0031 & 15114\\
  578 & PSZ1 G166.11+43.40 & RXC J0917.8+5143 & 09:17:53.55 & +51:43:42.3 & 0.217 & 7.0 & 0.0394 & 0.0024 & 5006\\
  582 & PSZ1 G167.64+17.63 & RXC J0638.1+4747 & 06:38:03.83 & +47:47:53.4 & 0.173 & 6.5 & 0.0334 & 0.0029 & 14388\\
  608 & PSZ1 G180.25+21.03 & RXC J0717.5+3745 & 07:17:31.97 & +37:45:28.7 & 0.546 & 11. & 0.0291 & 0.0017 & 1655,4200\\
  610 & PSZ1 G180.56+76.66 & RXC J1157.3+3336 & 11:57:17.39 & +33:36:38.4 & 0.213 & 6.0 & 0.1167 & 0.0038 & 11724,538\\
  617 & PSZ1 G182.55+55.83 & RXC J1017.0+3902 & 10:17:03.6 & +39:02:48.06  & 0.206 & 5.7 & 0.1049 & 0.0021 & 903\\
  628 & PSZ1 G186.37+37.26 & RXC J0842.9+3621 & 08:42:57.78 & +36:21:59.4 & 0.282 & 11. & 0.0349 & 0.0023 & 4217\\
  654 & PSZ1 G195.60+44.03 & RXC J0920.4+3030 & 09:20:26.83 & +30:29:36.2 & 0.295 & 6.3 & 0.0338 & 0.0033 & 15128,534\\
  655 & PSZ1 G195.78-24.29 & RXC J0454.1+0255 & 04:54:06.71 & +02:54:26.3 & 0.202 & 7.0 & 0.0300 & 0.0013 & 4215\\
  676 & PSZ1 G205.07-62.94 &   & 02:46:26.64 & -20:33:10.8 & 0.310 & 7.3 & 0.0189 & 0.0022 & 0674380501\\
  681 & PSZ1 G205.94-39.46 & RXC J0417.5-1154 & 04:17:34.74 & -11:54:34.0 & 0.442 & 11. & 0.1761 & 0.0076 & 3270\\
  686 & PSZ1 G208.59-26.00 & RXC J0510.7-0801 & 05:10:47.48 & -08:01:35.6 & 0.219 & 7.3 & 0.0495 & 0.0030 & 14011\\
  688 & PSZ1 G208.80-30.67 & RXC J0454.1-1014 & 04:54:06.90 & -10:13:18.7 & 0.247 & 6.9 & 0.0412 & 0.0015 & 12880,13190,430,901\\
  700 & PSZ1 G212.97-84.04 & RXC J0118.1-2658 & 01:18:11.08 & -26:57:57.4 & 0.227 & 6.1 & 0.0395 & 0.0030 & 9429\\
  715 & PSZ1 G216.60+47.00 & RXC J0949.8+1707 & 09:49:51.73 & +17:07:06.8 & 0.382 & 8.2 & 0.0547 & 0.0048 & 3274\\
  726 & PSZ1 G218.83+35.49 & RXC J0909.1+1059 & 09:09:12.72 & +10:58:27.9 & 0.175 & 5.5 & 0.0800 & 0.0029 & 924\\
  744 & PSZ1 G223.91-60.09 & RXC J0307.0-2840 & 03:07:01.98 & -28:39:56.2 & 0.253 & 6.7 & 0.1473 & 0.0053 & 9414\\
  758 & PSZ1 G226.19+76.78 & RXC J1155.3+2324 & 11:55:17.95 & +23:24:19.0 & 0.142 & 5.9 & 0.1023 & 0.0011 & 1661,5003,537\\
  759 & PSZ1 G226.19-21.92 & RXC J0552.8-2103 & 05:52:51.43 & -21:03:14.5 & 0.098 & 4.2 & 0.0474 & 0.0030 & 15307\\
  772 & PSZ1 G229.23-17.23 & RXC J0616.3-2156 & 06:16:24.73 & -21:56:15.6 & 0.171 & 5.9 & 0.0337 & 0.0030 & 15100\\
  773 & PSZ1 G229.70+77.97 & RXC J1201.3+2306 & 12:01:15.52 & +23:06:19.4 & 0.268 & 7.7 & 0.0236 & 0.0023 & 11762,16279\\
  774 & PSZ1 G229.92+15.28 & RXC J0817.4-0730 & 08:17:25.95 & -07:30:34.0 & 0.070 & 4.6 & 0.0743 & 0.0009 & 2211\\
  796 & PSZ1 G236.93-26.65 & RXC J0547.6-3152 & 05:47:36.60 & -31:52:07.1 & 0.148 & 5.2 & 0.0425 & 0.0021 & 9419\\
  801 & PSZ1 G239.29+24.75 & RXC J0909.1-0939 & 09:09:17.12 & -09:41:12.6 & 0.054 & 6.6 & 0.0321 & 0.0005 & 577\\
  802 & PSZ1 G239.30-26.01 & RXC J0553.4-3342 & 05:53:28.45 & -33:42:33.3 & 0.430 & 9.3 & 0.0424 & 0.0020 & 12244,5813\\
  815 & PSZ1 G241.75-30.89 & RXC J0532.9-3701 & 05:32:55.42 & -37:01:37.6 & 0.270 & 6.7 & 0.0770 & 0.0037 & 15112\\
  816 & PSZ1 G241.76-24.01 & RXC J0605.8-3518 & 06:05:53.95 & -35:18:07.7 & 0.139 & 5.4 & 0.2335 & 0.0065 & 15315\\
  818 & PSZ1 G241.98+14.87 & RXC J0841.9-1729 & 08:41:52.03 & -17:27:50.4 & 0.168 & 6.4 & 0.0385 & 0.0020 & 13378,15316\\
  822 & PSZ1 G243.14-73.87 & RXC J0159.0-3412 & 01:59:02.69 & -34:12:57.6 & 0.409 & 7.6 & 0.0300 & 0.0061 & 5818\\
  824 & PSZ1 G243.60+67.74 & RXC J1132.8+1428 & 11:32:51.92 & +14:27:11.4 & 0.083 & 4.1 & 0.0493 & 0.0028 & 14387\\
  826 & PSZ1 G244.35-32.15 & RXC J0528.9-3927 & 05:28:52.98 & -39:28:15.4 & 0.283 & 7.3 & 0.1139 & 0.0024 & 15177,15658,4994\\
  829 & PSZ1 G244.67+32.47 & RXC J0945.4-0839 & 09:45:27.04 & -08:39:24.9 & 0.153 & 5.0 & 0.0321 & 0.0027 & 15109\\
  838 & PSZ1 G246.53-26.07 & RXC J0601.7-3959 & 06:02:11.65 & -39:56:55.9 & 0.046 & 2.2 & 0.0279 & 0.0024 & 3202,3450\\
  840 & PSZ1 G247.19-23.31 & RXC J0616.5-3948 & 06:16:32.18 & -39:47:47.2 & 0.151 & 4.2 & 0.0530 & 0.0042 & 15126\\
  857 & PSZ1 G250.92-36.24 & RXC J0510.2-4519 & 05:10:17.06 & -45:19:10.8 & 0.200 & 5.9 & 0.0663 & 0.0043 & 15111\\
  862 & PSZ1 G252.99-56.06 & RXC J0317.9-4414 & 03:17:57.68 & -44:14:18.2 & 0.075 & 3.0 & 0.2612 & 0.0018 & 13135,6972,7323,7324\\
  868 & PSZ1 G253.49-33.73 & RXC J0525.8-4715 & 05:25:48.96 & -47:15:10.7 & 0.191 & 4.8 & 0.0690 & 0.0042 & 15122\\
  877 & PSZ1 G255.60-46.18 & SPT-CLJ0411-4819 & 04:11:16.40 & -48:18:53.8 & 0.423 & 6.8 & 0.0917 & 0.0047 & 13396,16355,17536\\
  880 & PSZ1 G256.55-65.69 & RXC J0225.9-4154 & 02:25:53.03 & -41:54:56.2 & 0.219 & 5.8 & 0.1338 & 0.0064 & 15110\\
  882 & PSZ1 G257.32-22.19 & RXC J0637.3-4828 & 06:37:14.66 & -48:28:18.2 & 0.202 & 4.8 & 0.0898 & 0.0051 & 15125\\
  889 & PSZ1 G260.00-63.45 & RXC J0232.2-4420 & 02:32:18.70 & -44:20:46.9 & 0.283 & 6.8 & 0.1423 & 0.0126 & 4993\\
  898 & PSZ1 G262.27-35.38 & RXC J0516.6-5430 & 05:16:36.65 & -54:30:49.5 & 0.295 & 9.0 & 0.0235 & 0.0028 & 15099,9331\\
  901 & PSZ1 G262.72-40.92 &  & 04:38:17.11 & -54:19:24.9 & 0.421 & 7.5 & 0.0791 & 0.0055 & 12259\\
  904 & PSZ1 G263.14-23.42 & RXC J0638.7-5358 & 06:38:48.53 & -53:58:26.6 & 0.226 & 6.7 & 0.1116 & 0.0035 & 9420\\
  905 & PSZ1 G263.19-25.22 & RXC J0627.2-5428 & 06:26:47.69 & -54:32:48.2 & 0.050 & 2.6 & 0.0529 & 0.0020 & 4944\\
  907 & PSZ1 G263.68-22.55 & RXC J0645.4-5413 & 06:45:28.62 & -54:13:40.7 & 0.164 & 7.8 & 0.0749 & 0.0037 & 15301\\
  912 & PSZ1 G264.62-51.07 & RXC J0330.8-5228 & 03:29:50.57 & -52:34:50.9 & 0.439 & 5.7 & 0.0296 & 0.0041 & 893\\
  914 & PSZ1 G265.02-48.96 & RXC J0342.8-5338 & 03:42:46.93 & -53:36:39.2 & 0.059 & 4.2 & 0.0276 & 0.0005 & 3201,3712\\
  920 & PSZ1 G266.02-21.23 & RXC J0658.5-5556 & 06:58:20.12 & -55:56:30.2 & 0.296 & 12. & 0.0568 & 0.0006 & 3184,4984,4985,4986,5355,5356,5357,5358,5361,554\\
  924 & PSZ1 G266.85+25.06 & RXC J1023.8-2715 & 10:23:50.25 & -27:15:21.6 & 0.254 & 7.6 & 0.2289 & 0.0030 & 9400\\
  931 & PSZ1 G269.28-49.89 & RXC J0328.6-5542 & 03:28:36.85 & -55:43:09.2 & 0.085 & 3.1 & 0.0638 & 0.0034 & 15313\\
  941 & PSZ1 G271.48-56.57 & ACO S 295 & 02:45:24.84 & -53:01:42.9 & 0.300 & 6.5 & 0.0451 & 0.0011 & 12260,16127,16282,16524,16525\\
  944 & PSZ1 G272.08-40.16 & RXC J0431.4-6126 & 04:31:13.25 & -61:27:14.0 & 0.058 & 6.7 & 0.0362 & 0.0007 & 899\\
  951 & PSZ1 G273.54+63.23 & RXC J1200.4+0320 & 12:00:24.74 & +03:20:37.9 & 0.133 & 5.6 & 0.0240 & 0.0019 & 15188,15306\\
  960 & PSZ1 G278.58+39.15 & RXC J1131.9-1955 & 11:31:54.42 & -19:55:42.3 & 0.307 & 8.8 & 0.0734 & 0.0058 & 3276\\
  971 & PSZ1 G280.21+47.83 & RXC J1149.7-1219 & 11:49:47.39 & -12:18:56.2 & 0.155 & 5.4 & 0.0293 & 0.0030 & 15311\\
  980 & PSZ1 G282.45+65.18 & RXC J1217.6+0339 & 12:17:41.06 & +03:39:18.4 & 0.076 & 4.6 & 0.0380 & 0.0015 & 4184\\
  984 & PSZ1 G284.43+52.44 & RXC J1206.2-0848 & 12:06:12.12 & -08:48:02.2 & 0.441 & 10. & 0.0858 & 0.0041 & 3277\\
  988 & PSZ1 G285.63+72.72 & RXC J1230.7+1033 & 12:30:47.58 & +10:33:11.9 & 0.165 & 5.6 & 0.0376 & 0.0029 & 12254\\
  993 & PSZ1 G286.27-38.39 &    & 03:59:09.12 & -72:04:33.6 & 0.307 & 6.0 & 0.0240 & 0.0033 & 0656200501\\
  994 & PSZ1 G286.60-31.23 &  & 05:31:28.81 & -75:10:36.2 & 0.209 & 5.2 & 0.0349 & 0.0034 & 15115\\
  1006 & PSZ1 G287.95-32.98 &    & 04:59:45.36 & -75:48:32.4 & 0.25 & 5.8 & 0.0278 & 0.0025 & 0762800101\\
  1009 & PSZ1 G288.26+39.94 & RXC J1203.2-2131 & 12:03:17.04 & -21:32:20.4 & 0.199 & 7.3 & 0.0251 & 0.0015 & 0652010101\\
  1011 & PSZ1 G288.63-37.67 & RXC J0352.4-7401 & 03:52:32.37 & -74:02:09.3 & 0.127 & 6.4 & 0.0375 & 0.0028 & 13380\\
  1032 & PSZ1 G294.68-37.01 & RXC J0303.7-7752 & 03:03:40.73 & -77:52:45.7 & 0.274 & 6.9 & 0.0380 & 0.0027 & 15113\\
  1037 & PSZ1 G295.34+23.34 & RXC J1215.4-3900 & 12:15:26.70 & -39:01:38.2 & 0.119 & 4.3 & 0.0245 & 0.0022 & 15140,15310\\
  1038 & PSZ1 G295.60-51.95 &    & 01:33:26.88 & -64:34:08.4 & 0.333 & 6.3 & 0.0439 & 0.0067 & 0762800301\\
  1041 & PSZ1 G296.42-32.49 & RXC J0351.1-8212 & 03:51:31.80 & -82:13:10.7 & 0.061 & 2.5 & 0.0524 & 0.0024 & 16283,8272\\
  1046 & PSZ1 G297.94-67.76 & SPT-CLJ0102-49151 & 01:02:58.27 & -49:16:27.1 & 0.870 & 8.7 & 0.1155 & 0.0023 & 12258,14022,14023\\
  1057 & PSZ1 G303.73+33.69 & RXC J1254.6-2913 & 12:54:40.64 & -29:13:40.0 & 0.054 & 3.2 & 0.1850 & 0.0050 & 8268\\
  1062 & PSZ1 G304.44+32.45 & RXC J1257.2-3022 & 12:57:21.98 & -30:21:47.7 & 0.055 & 3.0 & 0.0526 & 0.0025 & 10745\\
  1065 & PSZ1 G304.86-41.40 &  & 00:28:02.66 & -75:37:52.7 & 0.409 & 7.5 & 0.0340 & 0.0037 & 14390\\
  1066 & PSZ1 G304.91+45.46 & RXC J1257.1-1724 & 12:57:11.80 & -17:24:31.8 & 0.047 & 3.8 & 0.0784 & 0.0008 & 2206,7922\\
  1071 & PSZ1 G305.88-44.56 &      & 00:23:39.12 & -72:24:03.6 & 0.300 & 6.0 & 0.0400 & 0.0032 & 0679180301\\
  1078 & PSZ1 G306.71+61.04 & RXC J1258.6-0145 & 12:58:41.45 & -01:45:43.7 & 0.084 & 4.0 & 0.1110 & 0.0006 & 5822,5823,6356,6357,6358,7242\\
  1079 & PSZ1 G306.77+58.62 & RXC J1259.3-0411 & 12:59:22.16 & -04:11:50.1 & 0.084 & 5.1 & 0.0798 & 0.0021 & 4185\\
  1095 & PSZ1 G311.98+30.73 & RXC J1327.9-3130 & 13:27:56.88 & -31:29:45.6 & 0.048 & 4.4 & 0.0557 & 0.0004 & 0107260101\\
  1100 & PSZ1 G312.64+35.09 & RXC J1326.9-2710 & 13:26:58.24 & -27:10:55.2 & 0.045 & 2.9 & 0.0215 & 0.0010 & 4186\\
  1105 & PSZ1 G313.33+61.13 & RXC J1311.5-0120 & 13:11:29.52 & -01:20:24.4 & 0.183 & 8.8 & 0.1168 & 0.0015 & 1663,5004,540\\
  1109 & PSZ1 G313.88-17.12 & RXC J1601.7-7544 & 16:01:49.10 & -75:45:19.3 & 0.152 & 7.5 & 0.0814 & 0.0027 & 14386\\
  1117 & PSZ1 G315.69-18.05 & RXC J1631.6-7507 & 16:31:21.25 & -75:06:51.5 & 0.104 & 6.4 & 0.0269 & 0.0016 & 13377,15317\\
  1118 & PSZ1 G316.33+28.55 & RXC J1347.4-3250 & 13:47:28.32 & -32:51:57.6 & 0.039 & 4.6 & 0.0626 & 0.0004 & 0086950201\\
  1126 & PSZ1 G321.98-47.96 & RXC J2249.9-6425 & 22:49:56.87 & -64:25:48.3 & 0.093 & 4.3 & 0.0541 & 0.0015 & 4973\\
  1134 & PSZ1 G324.05+48.79 & RXC J1347.5-1144 & 13:47:30.60 & -11:45:09.6 & 0.451 & 10. & 0.2448 & 0.0026 & 3592,507\\
  1136 & PSZ1 G324.51-44.98 & RXC J2218.0-6511 & 22:18:00.10 & -65:10:52.5 & 0.095 & 3.5 & 0.1183 & 0.0054 & 15314\\
  1139 & PSZ1 G325.70+17.31 &  & 14:47:33.89 & -40:20:38.5 & 0.315 & 7.4 & 0.0373 & 0.0052 & 15298\\
  1157 & PSZ1 G332.21-46.38 & RXC J2201.9-5956 & 22:01:52.91 & -59:56:44.8 & 0.097 & 5.9 & 0.0536 & 0.0008 & 7920\\
  1160 & PSZ1 G332.87-19.26 & RXC J1813.3-6127 & 18:13:13.20 & -61:27:04.0 & 0.146 & 5.8 & 0.0459 & 0.0031 & 14389\\
  1164 & PSZ1 G335.57-46.47 & RXC J2154.1-5751 & 21:54:04.18 & -57:52:03.9 & 0.075 & 4.1 & 0.0435 & 0.0023 & 8269\\
  1165 & PSZ1 G336.61-55.43 & RXC J2246.3-5243 & 22:46:28.61 & -52:45:45.3 & 0.096 & 4.3 & 0.0293 & 0.0025 & 15304\\
  1182 & PSZ1 G340.37+60.57 & RXC J1401.0+0252 & 14:01:01.97 & +02:52:43.4 & 0.252 & 8.4 & 0.2624 & 0.0030 & 495\\
  1184 & PSZ1 G340.86-33.36 & RXC J2012.5-5649 & 20:12:42.07 & -56:50:48.8 & 0.055 & 5.7 & 0.0296 & 0.0002 & 513,5751,5752,5753,6292,6295,6296,889\\
  1185 & PSZ1 G340.94+35.10 & RXC J1459.4-1811 & 14:59:28.99 & -18:10:45.0 & 0.235 & 7.7 & 0.2723 & 0.0037 & 9428\\
  1190 & PSZ1 G342.33-34.92 & RXC J2023.4-5535 & 20:23:21.39 & -55:35:49.8 & 0.231 & 6.6 & 0.0351 & 0.0034 & 15108\\
  1192 & PSZ1 G342.83-30.47 & RXC J1952.2-5503 & 19:52:13.41 & -55:03:13.6 & 0.059 & 3.0 & 0.0279 & 0.0024 & 15308\\
  1200 & PSZ1 G346.61+35.06 & RXC J1514.9-1523 & 15:15:03.13 & -15:22:46.1 & 0.222 & 8.3 & 0.0163 & 0.0015 & 15175\\
  1201 & PSZ1 G347.20-27.36 & RXC J1934.7-5053 & 19:34:52.46 & -50:52:34.6 & 0.237 & 6.4 & 0.0286 & 0.0028 & 15120\\
  1207 & PSZ1 G348.92-67.38 & ACO S 1121 & 23:25:11.39 & -41:12:12.4 & 0.358 & 5.0 & 0.0623 & 0.0090 & 13405\\
  1208 & PSZ1 G349.46-59.92 & RXC J2248.7-4431 & 22:48:43.90 & -44:31:50.0 & 0.347 & 11. & 0.0625 & 0.0020 & 4966\\
  1214 & PSZ1 G352.35-77.66 & RXC J0006.0-3443 & 00:06:00.00 & -34:43:19.2 & 0.114 & 3.7 & 0.0273 & 0.0016 & 0201903801\\
  1216 & PSZ1 G355.07+46.20 & RXC J1504.1-0248 & 15:04:07.43 & -2:48:15.94 & 0.215 & 6.9 & 0.3392 & 0.0025 & 5793\\
  1218 & PSZ1 G356.18-76.06 & RXC J2357.0-3445 & 23:57:00.72 & -34:45:36.0 & 0.047 & 2.3 & 0.1284 & 0.0068 & 0109950201\\
  1224 & PSZ1 G358.96-67.26 & RXC J2315.7-3746 & 23:15:46.32 & -37:47:49.2 & 0.178 & 5.4 & 0.0480 & 0.0037 & 0762800501\\
\label{table:planck}
\end{longtable}
\end{footnotesize}
\endlandscape
\clearpage
\onecolumn
\begin{longtable}{c c c c c c c c}
\hline
 NAME & R.A.$_X$ & Dec.$_X$ & Redshift & $M_{500}$ & $c$ & $\sigma_c$ & Obs. ID \\
       &   &    &   & $10^{14} M_\odot$ &  &   &  \\
\hline      
   \endhead

\hline
\endfoot

\hline
\caption{Properties of the clusters in the ME-MACS sample. Col. [1] is the cluster name,  Cols. [2] and [3] are the coordinates of the X-ray peak that we used to measure the concentration parameter, col. [4] is the redshift of the cluters and col[5]is its mass calculated from the X-ray luminosity in \citep{mann_ebe}, col. [6] and col. [7] are the concentration parameter  and its error, while in col [8] we list the ID of the observations  used in our analysis (those starting with ``0'' are \xmm\ data). }
\endlastfoot

A1914 & 14:26:03.12 & +37:49:24.9 & 0.17 & 8.08 & 0.0626 & 0.0014 & 3593\\
  A209 & 01:31:52.76 & -13:36:41.4 & 0.21 & 5.71 & 0.0482 & 0.0025 & 3579 522\\
  A586 & 07:32:20.61 & +31:37:49.4 & 0.17 & 5.99 & 0.0835 & 0.0028 & 11723 530\\
  ABELL1689 & 13:11:29.52 & -01:20:24.4 & 0.18 & 9.46 & 0.1188 & 0.0016 & 6930\\
  ABELL1758 & 13:32:38.49 & +50:33:35.0 & 0.28 & 5.41 & 0.0263 & 0.0016 & 2213\\
  ABELL1835 & 14:01:02.10 & +02:52:45.4 & 0.25 & 11.8 & 0.2500 & 0.0029 & 6880\\
  ABELL2163 & 16:15:46.06 & -06:09:06.0 & 0.21 & 12.3 & 0.0224 & 0.0006 & 1653\\
  ABELL2204 & 16:32:46.94 & +05:34:32.1 & 0.15 & 9.60 & 0.3051 & 0.0013 & 7940\\
  ABELL2219 & 16:40:21.46 & +46:42:27.4 & 0.23 & 8.78 & 0.0283 & 0.0007 & 7892\\
  ABELL2261 & 17:22:27.30 & +32:07:55.4 & 0.22 & 7.58 & 0.1143 & 0.0030 & 5007\\
  ABELL2390 & 21:53:36.81 & +17:41:45.2 & 0.23 & 9.22 & 0.1249 & 0.0011 & 4193\\
  ABELL2631 & 23:37:38.08 & +00:16:02.2 & 0.28 & 6.41 & 0.0308 & 0.0039 & 11728 3248\\
  ABELL2667 & 23:51:39.37 & -26:05:01.8 & 0.23 & 9.18 & 0.1648 & 0.0040 & 2214\\
  ABELL2744 & 00:14:19.04 & -30:23:30.0 & 0.31 & 9.35 & 0.0223 & 0.0009 & 8477 8557\\
  ABELL2813 & 00:43:24.37 & -20:37:33.8 & 0.29 & 5.83 & 0.0489 & 0.0032 & 9409\\
  ABELL3088 & 03:07:01.82 & -28:39:56.4 & 0.25 & 5.49 & 0.1389 & 0.0051 & 9414\\
  ABELL520 & 04:54:05.11 & +02:53:37.0 & 0.2 & 6.86 & 0.0305 & 0.0016 & 4215\\
  ABELL521 & 04:54:06.56 & -10:13:14.9 & 0.25 & 6.38 & 0.0391 & 0.0023 & 901\\
  ABELL611 & 08:00:56.83 & +36:03:23.4 & 0.29 & 5.56 & 0.1232 & 0.0036 & 3194\\
  ABELL665 & 08:30:58.52 & +65:50:24.7 & 0.18 & 7.42 & 0.0485 & 0.0018 & 3586\\
  ABELL697 & 08:42:57.60 & +36:21:59.4 & 0.28 & 7.74 & 0.0350 & 0.0022 & 4217\\
  ABELL773 & 09:17:51.21 & +51:43:38.3 & 0.22 & 5.97 & 0.0385 & 0.0023 & 5006 533\\
  Abell1682 & 13:06:49.79 & +46:33:29.8 & 0.23 & 5.77 & 0.0427 & 0.0034 & 11725 3244\\
  Abell2146 & 15:56:14.83 & +66:20:49.1 & 0.23 & 5.68 & 0.1215 & 0.0009 & 10464\\
  Abell2552 & 23:11:33.25 & +03:38:08.2 & 0.3 & 7.04 & 0.0629 & 0.0032 & 11730\\
  IRAS09104+4109 & 09:13:45.57 & +40:56:28.9 & 0.44 & 6.62 & 0.2998 & 0.0056 & 10445\\
  MACS-J0111.5+0855 & 01:11:31.60 & +08:55:41.9 & 0.49 & 6.22 & 0.1609 & 0.0167 & 3256\\
  MACS-J2129.4-0741 & 21:29:25.68 & -07:41:23.7 & 0.59 & 6.04 & 0.0576 & 0.0051 & 3199 3595\\
  MACS-J2243.3-0935 & 22:43:21.19 & -09:35:37.2 & 0.45 & 8.14 & 0.0238 & 0.0032 & 3260\\
  MACS1108.8+0906 & 11:08:55.82 & +09:05:53.6 & 0.47 & 5.07 & 0.0394 & 0.0041 & 5009\\
  MACS1427+44 & 14:27:16.09 & +44:07:30.3 & 0.49 & 6.92 & 0.2531 & 0.0070 & 9380\\
  MACS1427-25 & 14:27:39.44 & -25:21:01.6 & 0.32 & 6.45 & 0.2735 & 0.0076 & 3279 9373\\
  MACS911.2+1746 & 09:11:10.90 & +17:46:28.9 & 0.51 & 5.55 & 0.0381 & 0.0044 & 3587 5012\\
  MACSJ0011.7-1523 & 00:11:42.89 & -15:23:20.6 & 0.38 & 6.41 & 0.1532 & 0.0046 & 3261 6105\\
  MACSJ0025.4-1222 & 00:25:29.43 & -12:22:40.5 & 0.58 & 6.42 & 0.0309 & 0.0019 & 10413\\
  MACSJ0035.4-2015 & 00:35:26.31 & -20:15:46.8 & 0.35 & 6.02 & 0.0519 & 0.0032 & 3262\\
  MACSJ0140.0-0555 & 01:40:01.49 & -05:55:11.0 & 0.45 & 6.54 & 0.0503 & 0.0049 & 5013 12243\\
  MACSJ0152.5-2852 & 01:52:34.45 & -28:53:37.8 & 0.41 & 7.66 & 0.1112 & 0.0078 & 3264\\
  MACSJ0159.0-3412 & 01:59:04.00 & -34:12:49.7 & 0.41 & 5.53 & 0.0212 & 0.0055 & 5818\\
  MACSJ0159.8-0849 & 01:59:49.30 & -08:49:58.5 & 0.41 & 7.56 & 0.2194 & 0.0045 & 6106\\
  MACSJ0242.5-2132 & 02:42:35.99 & -21:32:26.6 & 0.31 & 7.67 & 0.3821 & 0.0111 & 3266\\
  MACSJ0257.1-2325 & 02:57:08.73 & -23:26:08.0 & 0.51 & 6.09 & 0.0935 & 0.0049 & 1654 3581\\
  MACSJ0257.6-2209 & 02:57:40.97 & -22:09:17.4 & 0.32 & 5.74 & 0.1095 & 0.0058 & 3267\\
  MACSJ0308.9+2645 & 03:08:56.06 & +26:45:42.5 & 0.36 & 6.01 & 0.0514 & 0.0031 & 3268\\
  MACSJ0326.8-0043 & 03:26:49.88 & -00:43:51.8 & 0.45 & 6.43 & 0.3432 & 0.0162 & 5810\\
  MACSJ0329.6-0211 & 03:29:41.60 & -02:11:45.5 & 0.45 & 5.94 & 0.2747 & 0.0060 & 6108\\
  MACSJ0358.8-2955 & 03:58:54.30 & -29:55:33.3 & 0.43 & 8.48 & 0.0837 & 0.0025 & 12300 13194 11719\\
  MACSJ0404.6+1109 & 04:04:33.46 & +11:08:00.0 & 0.36 & 6.44 & 0.0425 & 0.0064 & 3269\\
  MACSJ0416.1-2403 & 04:16:09.22 & -24:04:02.3 & 0.4 & 6.20 & 0.0491 & 0.0010 & 10446\\
  MACSJ0417.5-1154 & 04:17:34.60 & -11:54:30.0 & 0.44 & 11.0 & 0.1571 & 0.0070 & 11759\\
  MACSJ0429.6-0253 & 04:29:35.97 & -02:53:08.1 & 0.4 & 6.78 & 0.2924 & 0.0097 & 3271\\
  MACSJ0451.9+0006 & 04:51:54.39 & +00:06:19.4 & 0.43 & 5.04 & 0.0638 & 0.0092 & 5815\\
  MACSJ0455.2+0657 & 04:55:17.11 & +06:57:47.7 & 0.45 & 7.00 & 0.0997 & 0.0104 & 5812\\
  MACSJ0520.7-1328 & 05:20:41.93 & -13:28:49.7 & 0.34 & 6.39 & 0.0951 & 0.0055 & 3272\\
  MACSJ0547.0-3904 & 05:47:01.54 & -39:04:27.0 & 0.32 & 5.51 & 0.2963 & 0.0106 & 3273\\
  MACSJ0553.4-3342 & 05:53:28.45 & -33:42:33.3 & 0.43 & 6.08 & 0.0351 & 0.0018 & 5813\\
  MACSJ0717.5+3745 & 07:17:31.97 & +37:45:28.7 & 0.55 & 8.52 & 0.0277 & 0.0018 & 4200\\
  MACSJ0744.8+3927 & 07:44:52.80 & +39:27:26.0 & 0.7 & 7.52 & 0.1663 & 0.0049 & 6111\\
  MACSJ0949.8+1708 & 09:49:51.73 & +17:07:06.8 & 0.38 & 8.37 & 0.0533 & 0.0048 & 3274\\
  MACSJ1006.9+3200 & 10:06:54.90 & +32:01:33.0 & 0.4 & 5.08 & 0.0423 & 0.0078 & 5819\\
  MACSJ1105.7-1014 & 11:05:46.01 & -10:14:36.6 & 0.41 & 5.29 & 0.0468 & 0.0074 & 5817\\
  MACSJ1115.2+5320 & 11:15:14.89 & +53:19:56.1 & 0.47 & 5.03 & 0.0521 & 0.0037 & 5008\\
  MACSJ1115.8+0129 & 11:15:51.87 & +01:29:56.2 & 0.35 & 7.49 & 0.1993 & 0.0041 & 9375\\
  MACSJ1131.8-1955 & 11:31:54.84 & -19:55:46.2 & 0.31 & 6.71 & 0.0664 & 0.0054 & 3276\\
  MACSJ1149.5+2223 & 11:49:35.28 & +22:24:04.8 & 0.54 & 6.43 & 0.0316 & 0.0032 & 1656 3589\\
  MACSJ1206.2-0847 & 12:06:12.12 & -08:48:02.3 & 0.44 & 7.19 & 0.0856 & 0.0041 & 3277\\
  MACSJ1226.8+2153 & 12:26:51.05 & +21:49:50.0 & 0.44 & 4.79 & 0.0932 & 0.0036 & 3590\\
  MACSJ1311.0-0310 & 13:11:01.45 & -03:10:39.8 & 0.49 & 6.19 & 0.1670 & 0.0052 & 6110\\
  MACSJ1319.9+7003 & 13:20:08.12 & +70:04:37.0 & 0.33 & 5.86 & 0.0883 & 0.0089 & 3278\\
  MACSJ1359.1-1929 & 13:59:10.30 & -19:29:25.4 & 0.45 & 5.31 & 0.2751 & 0.0088 & 9378\\
  MACSJ1621.3+3810 & 16:21:24.87 & +38:10:07.7 & 0.46 & 5.68 & 0.2339 & 0.0060 & 6109\\
  MACSJ1731.6+2252 & 17:31:38.21 & +22:51:49.3 & 0.39 & 7.15 & 0.0154 & 0.0031 & 3281\\
  MACSJ1931.8-2634 & 19:31:49.59 & -26:34:34.3 & 0.35 & 8.41 & 0.3017 & 0.0081 & 9382\\
  MACSJ2046.0-3430 & 20:46:00.55 & -34:30:15.3 & 0.42 & 5.49 & 0.3067 & 0.0077 & 9377\\
  MACSJ2049.9-3217 & 20:49:55.25 & -32:16:51.6 & 0.32 & 5.42 & 0.0457 & 0.0038 & 3283\\
  MACSJ2211.7-0349 & 22:11:46.13 & -03:49:47.5 & 0.4 & 8.01 & 0.0970 & 0.0041 & 3284\\
  MACSJ2214.9-1359 & 22:14:57.13 & -14:00:16.6 & 0.5 & 6.52 & 0.0502 & 0.0039 & 3259 5011\\
  MACSJ2228+2036 & 22:28:33.72 & +20:37:16.6 & 0.41 & 7.47 & 0.0556 & 0.0048 & 3285\\
  MACSJ2229.7-2755 & 22:29:45.18 & -27:55:37.8 & 0.32 & 6.33 & 0.3643 & 0.0114 & 3286 9374\\
  MACSJ2245.0+2637 & 22:45:04.48 & +26:38:03.3 & 0.3 & 6.26 & 0.1846 & 0.0084 & 3287\\
  MS0016.9+1609 & 00:18:33.75 & +16:26:11.6 & 0.55 & 6.73 & 0.0380 & 0.0022 & 520\\
  MS0451.6-0305 & 04:54:10.93 & -03:00:55.9 & 0.54 & 7.60 & 0.0429 & 0.0023 & 529\\
  MS0735.6+7421 & 07:41:44.13 & +74:14:39.9 & 0.22 & 5.72 & 0.1866 & 0.0031 & 10470\\
  MS1455.0+2232 & 14:57:15.07 & +22:20:32.7 & 0.26 & 7.10 & 0.2832 & 0.0029 & 4192\\
  MS2137.3-2353 & 21:40:15.17 & -23:39:39.7 & 0.31 & 6.50 & 0.3503 & 0.0111 & 4974\\
  RBS0797 & 09:47:12.86 & +76:23:13.1 & 0.35 & 8.90 & 0.3189 & 0.0035 & 2202\\
  RCS1447+0828 & 14:47:25.87 & +08:28:24.8 & 0.38 & 9.73 & 0.3132 & 0.0070 & 10481\\
  RXCJ0528.9-3927 & 05:28:52.98 & -39:28:15.4 & 0.28 & 8.22 & 0.1139 & 0.0024 & 4994\\
  RXCJ1023.8-2715 & 10:23:50.25 & -27:15:21.6 & 0.25 & 8.68 & 0.2173 & 0.0029 & 9400\\
  RXCJ1459.4-1811 & 14:59:28.99 & -18:10:45.0 & 0.23 & 7.10 & 0.2551 & 0.0036 & 9428\\
  RXJ0437.1+0043 & 04:37:09.47 & +00:43:55.6 & 0.28 & 6.14 & 0.1481 & 0.0038 & 11729\\
  RXJ0439.0+0715 & 04:39:00.37 & +07:16:06.1 & 0.24 & 5.48 & 0.0993 & 0.0035 & 3583\\
  RXJ0647.7+7015 & 06:47:50.88 & +70:14:50.6 & 0.59 & 7.20 & 0.0626 & 0.0043 & 3196 3584\\
  RXJ1347.5-1145 & 13:47:30.60 & -11:45:09.6 & 0.45 & 13.5 & 0.2421 & 0.0026 & 3592\\
  RXJ1423.8+2404 & 14:23:47.96 & +24:04:43.8 & 0.54 & 5.90 & 0.3159 & 0.0045 & 1657\\
  RXJ1504.1-0248 & 15:04:07.69 & -02:48:15.9 & 0.22 & 13.9 & 0.3196 & 0.0024 & 5793\\
  RXJ1532.9+3021 & 15:32:53.89 & +30:20:59.8 & 0.36 & 8.05 & 0.2442 & 0.0052 & 1665\\
  RXJ1720.1+2638 & 17:20:09.92 & +26:37:30.9 & 0.16 & 7.12 & 0.2320 & 0.0033 & 3224 4361\\
  RXJ1720.2+3536 & 17:20:16.69 & +35:36:26.1 & 0.39 & 6.94 & 0.2244 & 0.0052 & 3280 6107\\
  RXJ2014.8-2430 & 20:14:51.49 & -24:30:22.8 & 0.15 & 7.90 & 0.3286 & 0.0034 & 11757\\
  RXJ2129.6+0005 & 21:29:39.99 & +00:05:20.1 & 0.23 & 7.81 & 0.2116 & 0.0070 & 9370\\
  Z7215 & 15:01:22.73 & +42:20:44.7 & 0.29 & 5.37 & 0.0390 & 0.0046 & 7899\\
  ZWCL3146 & 10:23:39.60 & +04:11:12.1 & 0.29 & 11.4 & 0.2213 & 0.0027 & 909 9371\\
  ZwCl0348 & 01:06:49.44 & +01:03:23.0 & 0.25 & 5.64 & 0.3577 & 0.0048 & 10465\\

\label{table:macs}
\end{longtable}




\bsp	
\label{lastpage}
\end{document}